# SiC-BN composites with extreme thermal and environmental stability

Abhijit Biswas[1]*, Eugenio Garcia-Granados[2], Saman Zare[3], Mozhdeh Mirakhory[4], JungHun Park[5], Shuvodeep De[6], Jessica Reyes[3], Hunter B. Schonfeld[3], Noah A. Burke[7], William Riffe[7], Michael S. Lafferty[3], Sudaice Kazibwe[8], Maryam Amiri[1], Shreyasi Chattopadhyay[1], Mingrui Xu[1], Gelu Costin[9], Tanguy Terlier[10,11], Jianhua Li[12], Ching-Wu Chu[8], Liangzi Deng[8], Bharat K. Jasthi[13], Manoj Tripathi[14], Venkataramana Gadhamshetty[14], Gi-Dong Sim[5], Kathy Lu[15], Adri C. T. van Duin[4,16,17], Patrick E. Hopkins[3,7,18]*, Sanjay Sampath[2]* and Pulickel M. Ajayan[1]*

**AFFILIATIONS**

[1]Department of Materials Science and Nanoengineering, Rice University, Houston, TX, 77005, USA

[2]Center for Thermal Spray Research, Stony Brook University, 130 Heavy Engineering Bldg., Stony Brook, NY 11794-2275, USA

[3]Department of Mechanical and Aerospace Engineering, University of Virginia, Charlottesville, VA 22904, USA

[4]Department of Mechanical Engineering, The Pennsylvania State University, University Park, PA 16802, USA

[5]Department of Mechanical Engineering, Korea Advanced Institute of Science and Technology, 291, Daehak-ro, Yuseong-gu, Daejeon 34141, Republic of Korea

[6]Ingram School of Engineering, Texas State University, San Marcos, TX, 78666-4684, USA

[7]Department of Materials Science and Engineering, University of Virginia, Charlottesville, VA 22904, USA

[8]Department of Physics and Texas Center for Superconductivity (TcSUH), University of Houston, Houston, TX, 77004, USA

[9]Department of Earth, Environmental and Planetary Sciences, Rice University, Houston, TX, 77005, USA

[10]SIMS Laboratory, Shared Equipment Authority, Rice University, Houston, TX, 77005, USA

[11]Department of Chemical & Biomolecular Engineering, Rice University, Houston, TX, 77005, USA

[12]Shared Equipment Authority, Rice University, Houston, TX, 77005, USA

[13]Department of Materials and Metallurgical Engineering, South Dakota School of Mines & Technology, 501 East Saint Joseph St, Rapid City, SD 57701, USA

[14]Civil & Environmental Engineering, South Dakota School of Mines & Technology, Rapid City, SD 57701, USA

[15]Department of Mechanical and Materials Engineering, University of Alabama at Birmingham, Birmingham, AL, 35294, USA

[16]Department of Chemical Engineering, The Pennsylvania State University, University Park, PA 16802, USA

[17]Department of Materials Science and Engineering, The Pennsylvania State University, University Park, PA 16802, USA

[18]Department of Physics, University of Virginia, Charlottesville, VA 22904, USA

These authors contributed equally: Eugenio Garcia-Granados, Saman Zare, Mozhdeh Mirakhory, JungHun Park, Shuvodeep De

*Corresponding authors: **abhijit.biswas@rice.edu; phopkins@virginia.edu; sanjay.sampath@stonybrook.edu; ajayan@rice.edu**

## Graphical Abstract

Lightweight, machinable SiC-BN composites exhibiting unusually stable temperature-independent thermal conductivity and remarkable extreme environment stability, would be promising as a structural material in extreme conditions.

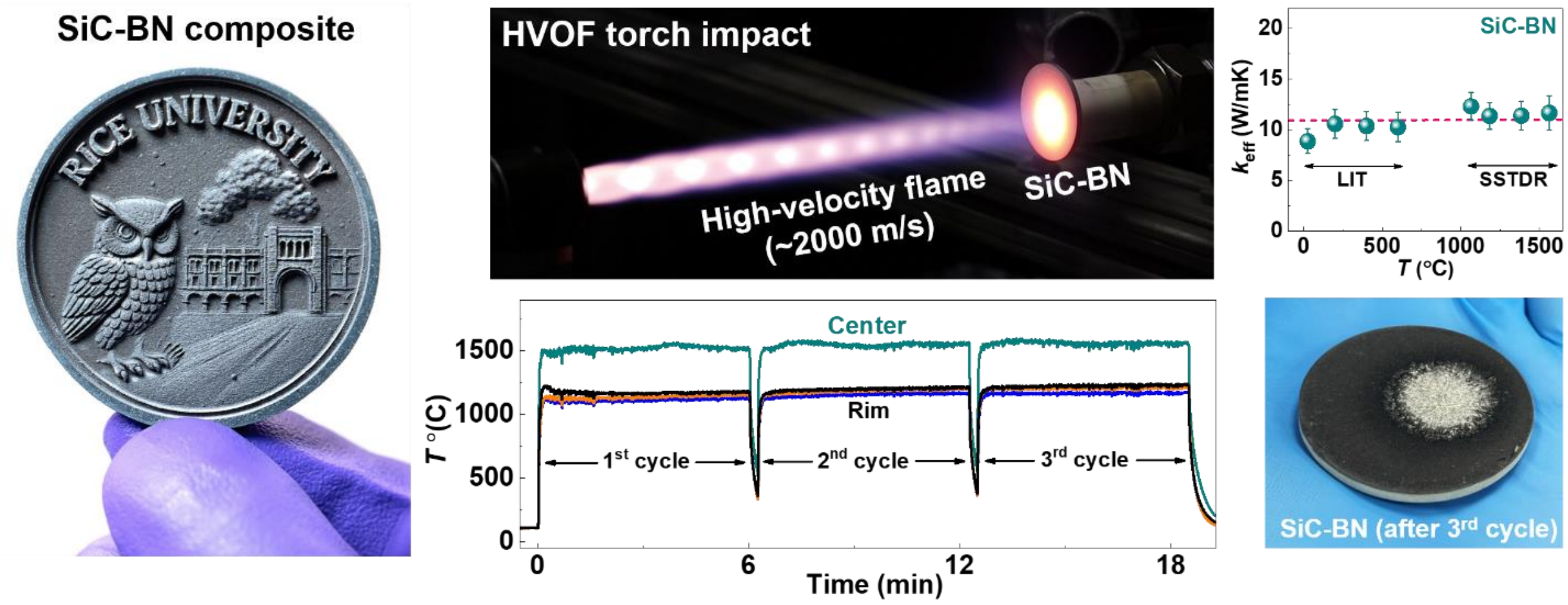

## Abstract

Ceramic materials that maintain structural integrity and resist thermal-shock under extreme conditions are critical for many applications. Here we show that spark plasma sintered composites containing equal weight fractions of hexagonal silicon carbide (6H-SiC) and boron nitride (h-BN) exhibit exceptional stability under extreme high-temperature oxidizing conditions and unusually stable temperature-independent thermal conductivity across a broad temperature range. Under high-velocity oxygen-fuel torch exposure in air, it withstands prolonged heating and thermal-cycling at surface temperatures reaching >1500 °C, whereas pure 6H-SiC and h-BN rapidly fractures or erodes. Microstructural evaluation shows the formation of a multi-phase surface silica layer that further protects the material from oxidation. Finite-element analysis heat-flux map shows that intrinsic porosity mitigates thermal stress, while ReaxFF molecular dynamics simulations reveal thermally driven interfacial reconstruction and the formation of a Si-N/C-B-rich bonding network, enhancing the interfacial cohesion at elevated temperatures. This lightweight, easily machinable ceramic composite will be a promising structural material in extreme environment applications.

## Main

Ceramics capable of maintaining structural integrity under extreme mechanical, thermal, and oxidative environments are essential for applications, ranging from space exploration to energy and industrial processing[1–6]. Monolithic ultrahigh-temperature ceramics are key to many high-performance systems owing to their chemical stability ≥2000 °C, yet their ability to endure such extreme conditions remains a persistent limitation[7–9]. Their high-density, brittleness with limited machinability, restricts their use in complex, and weight-sensitive designs. Furthermore, steep thermal gradients, chemically aggressive environments, and high mechanical loading induces thermal stress, promoting surface ablation and microstructural degradation, ultimately leading to catastrophic failures[7,10].

Composite ceramic materials offer a compelling pathway to overcome these limitations by integrating multiple phases across length scales, as they can decouple and redistribute thermal and mechanical loads, enabling enhanced thermal-shock resistance, oxidation, and erosion. In composites, tailored architectures, such as porous networks, and/or layered interfaces can promote energy dissipation and crack deflection, thereby mitigating catastrophic failures[1,11–14]. In applications, e.g. hypersonic vehicle skins and leading edges, propulsion components, and advanced manufacturing, materials are not required to endure the absolute highest temperatures, but rather to reliably survive application-relevant thermal conditions, while offering structural integrity, low-density, and ease of fabrication[15–18]. Therefore, lightweight ceramic composites that can address this critical gap by enabling precise manufacturing with design flexibility, stable thermal transport with excellent thermal-shock resistance, would be more suitable as extreme condition materials where efficiency and adaptability are as important as thermal endurance[18–20].

Here, we report the production of spark plasma sintered ceramic composite comprising equal wt% of hexagonal silicon carbide (6H-SiC) and boron nitride (h-BN) powders. This lightweight (density ~1.85 g/cm$^3$) and machinable SiC-BN composite exhibits unusual temperature-independent thermal conductivity from 25-1600 °C. During prolonged high-velocity oxygen-fuel (HVOF) torch exposure as well as thermal-cycling, the composite retains its structure at ≥1500 °C, whereas pure 6H-SiC and h-BN undergo catastrophic fracture or rapid erosion. Combined microstructural evaluation, finite element analyses, and temperature-dependent ReaxFF molecular dynamics simulations show: (a) formation of a mixed-phase surface silica layer, (b) porosity mitigating thermal stress, and (c) thermally-driven interfacial reconstruction with the formation of a Si-N/C-B-rich bonding network, contributing synergistically to its structural integrity. This shows the potential of SiC-BN composite as a robust structural material in extreme environments.

## Results

## Microstructural characterizations of SiC-BN composites

The SiC-BN ceramic composites were fabricated (in 50-50 wt%) via spark plasma sintering (SPS) by using 6H-SiC (particle size: 100 nm) and h-BN (particle size: 800 nm) powders (see **methods** and Supplementary information **Fig. 1**). After sintering, we obtained a light black colored disk (inset **Fig. 1a**) with a density of ~1.85±0.02 g/cm$^3$. We performed X-ray diffraction (XRD) upto 1000 °C, showing both the 6H-SiC (JCPDS: 00-022-1273) and h-BN (JCPDS: 00-009-0012) phases and their phase stability (**Fig. 1a** and Supplementary **Fig. 2**). High precision complex designs of the disk are made without any crack/fracture (inset of **Fig. 1b** and Supplementary **Fig. 3**)[21,22]. Thermal oxidation-resistance behavior shows a slight increase in mass above 900 °C (**Fig. 1b)**[23]. Electrical resistivity shows semiconducting behavior (**Fig. 1c)**[24]. X-ray microscopy (XRM) reveals pores (pore density ~1.65%) in the bulk (**Fig. 1d**). Cross-sectional field emission scanning electron microscopy (FESEM) shows both grains, and also the pores (**Fig. 1e**)[25]. Electron probe micro-analysis (EPMA) backscattered electron (BSE) image and its wavelength dispersive spectrometry (WDS) compositional map (Supplementary **Fig. 4**) confirms the presence of elements with an obtained wt% of 32.93% (Si), 24.38% (C), 12.56% (B), 27.89 %(N), and 2.34 %(O). High-resolution transmission electron microscopy (HRTEM) image shows both 6H-SiC and h-BN regions (forming an interface) (**Fig. 1f**), as confirmed by the diffraction patterns from respective regions (**Fig. 1g,h**), with Moiré patterns from h-BN multi-layers (twist angles $\theta_1$ ~9.2° and $\theta_2$ ~19.1°) (**Fig. 1h** and Supplementary **Fig. 5**). Although the interface is not atomically smooth, the reactive force-field (ReaxFF)[26,27] molecular dynamics simulations reveals that at ambient conditions Si-on-N is the most stable interface with the lowest formation energy (Supplementary **Fig. 6** and Supplementary **Table 1**). X-ray photoelectron spectroscopy (XPS) bonding analysis shows formation of these interfacial linkage, including Si-N (102.44 eV) and C-B (189.22 eV) bond's along with Si-C (100.54 eV) and B-N (190.98 eV) bonds from the constituent phases (**Fig. 1i, j** and Supplementary **Fig. 7**).

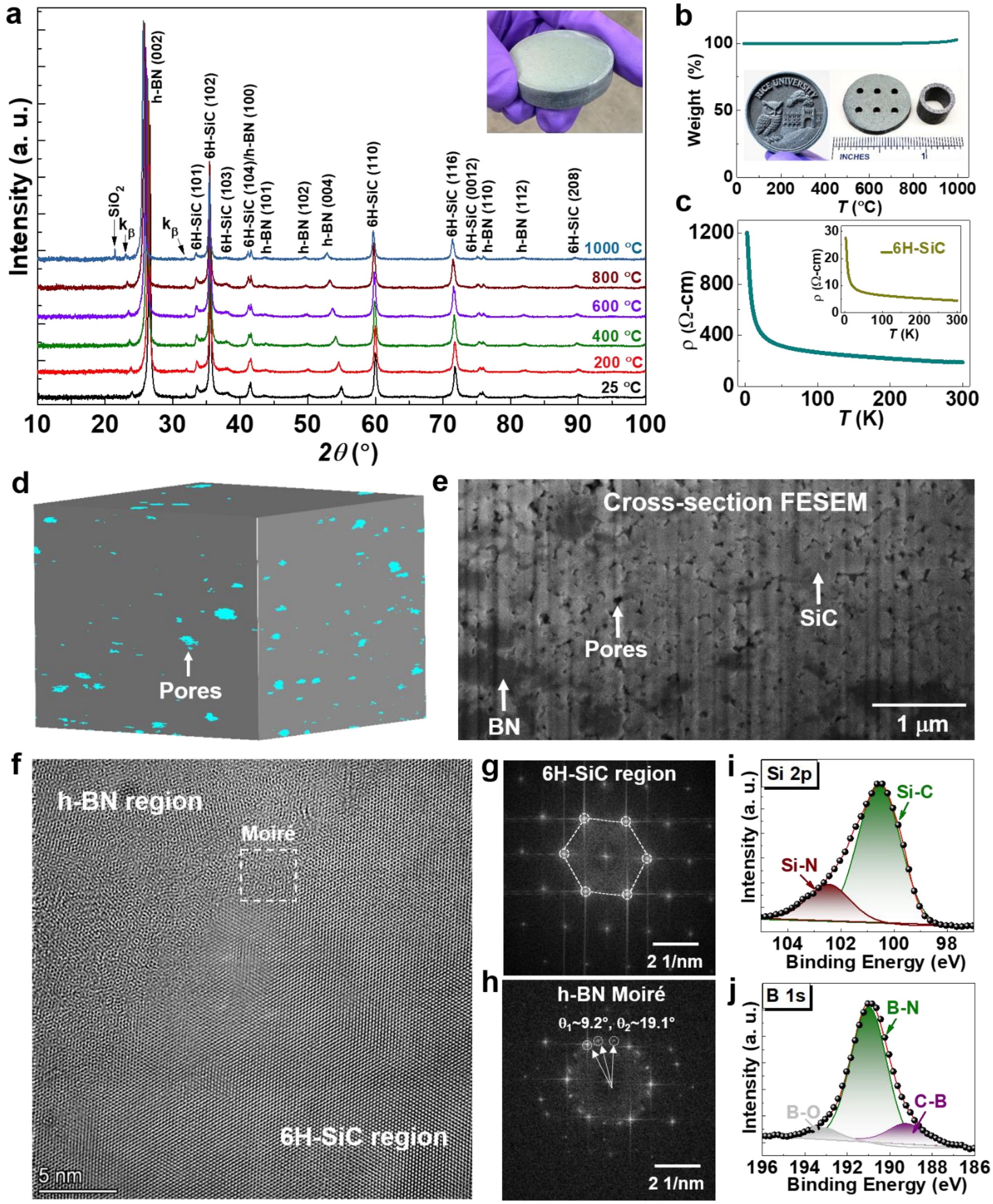

**Fig. 1: Characterization of sintered SiC-BN composites. a,** Temperature-dependent XRD pattern in air showing the Bragg reflections of 6H-SiC and h-BN phase, respectively, demonstrating the phase stability. The inset shows a sintered light black disk (40 mm diameter and 10 mm thickness). **b,** Oxidation-resistance behavior of the composite. The inset shows high-precision patterning or shapes of the composite disk, demonstrating its machinability. **c,** Electrical resistivity exhibiting semiconducting behavior. The inset shows the resistivity of SPS 6H-SiC. **d,** X-ray microscopy image revealing pores in the bulk (cyan: pores, gray: SiC-BN matrix). Box dimension: $1.35\times1.35\times1.10$ $mm^3$. **e,** Cross-sectional FESEM showing the microstructure. **f,** HRTEM image shows h-BN and 6H-SiC regions with an interface. **g, h,** Corresponding diffraction patterns show hexagon from 6H-SiC and Moiré from layered h-BN with twist angles ~9.2° and ~19.1°. **i, j,** Element-resolved XPS spectra shows Si-C and B-N bonds within the constituent phases, together with Si-N and C-B bonding, indicating the interfacial linkages.

## Mechanical properties and thermal conductivity

To evaluate the mechanical properties, nanoindentation was conducted at 25 °C and 800 °C (Supplementary **Fig. 8** and **9**). At 25 °C, the composite exhibits a Young's modulus ($E$) of 35.5 ± 3.6 and hardness ($H$) of 1.38 ± 0.24 GPa (**Fig. 2a**). These properties were largely retained at 800 °C with $E$ and $H$ values of 34.1 ± 3.1 GPa and $H$ =1.34 ± 0.13 GPa, respectively, demonstrating the composite's high-temperature mechanical stability[28]. Nanoindentation-based analysis yielded a fracture toughness ($K_c$) of 2.02 ± 0.80 MPa·$m^{1/2}$. Micropillar compression tests conducted at room temperature (Supplementary **Fig. 10** and **Table 2**) gave a yield strength ($\sigma_y$) and a failure strength of 0.86 ± 0.23 and 0.89 ± 0.20 GPa, respectively. The Young's modulus obtained from micropillar compression tests ($E$ =30 ± 7 GPa) was comparable to that measured by nanoindentation tests. In addition, the tribological linear wear tests were also conducted against zirconia balls. The wear map shows a steadily increasing friction, with the broad and irregular wear-scars, characterized by the severe ploughing and debris accumulation. Under reciprocating wear conditions, the obtained wear rate at 25 °C is ~0.234 $mm^3$/N·m (at 1 N) and ~0.224 $mm^3$/N·m (at 5 N), respectively (Supplementary **Fig. 11**)[29].

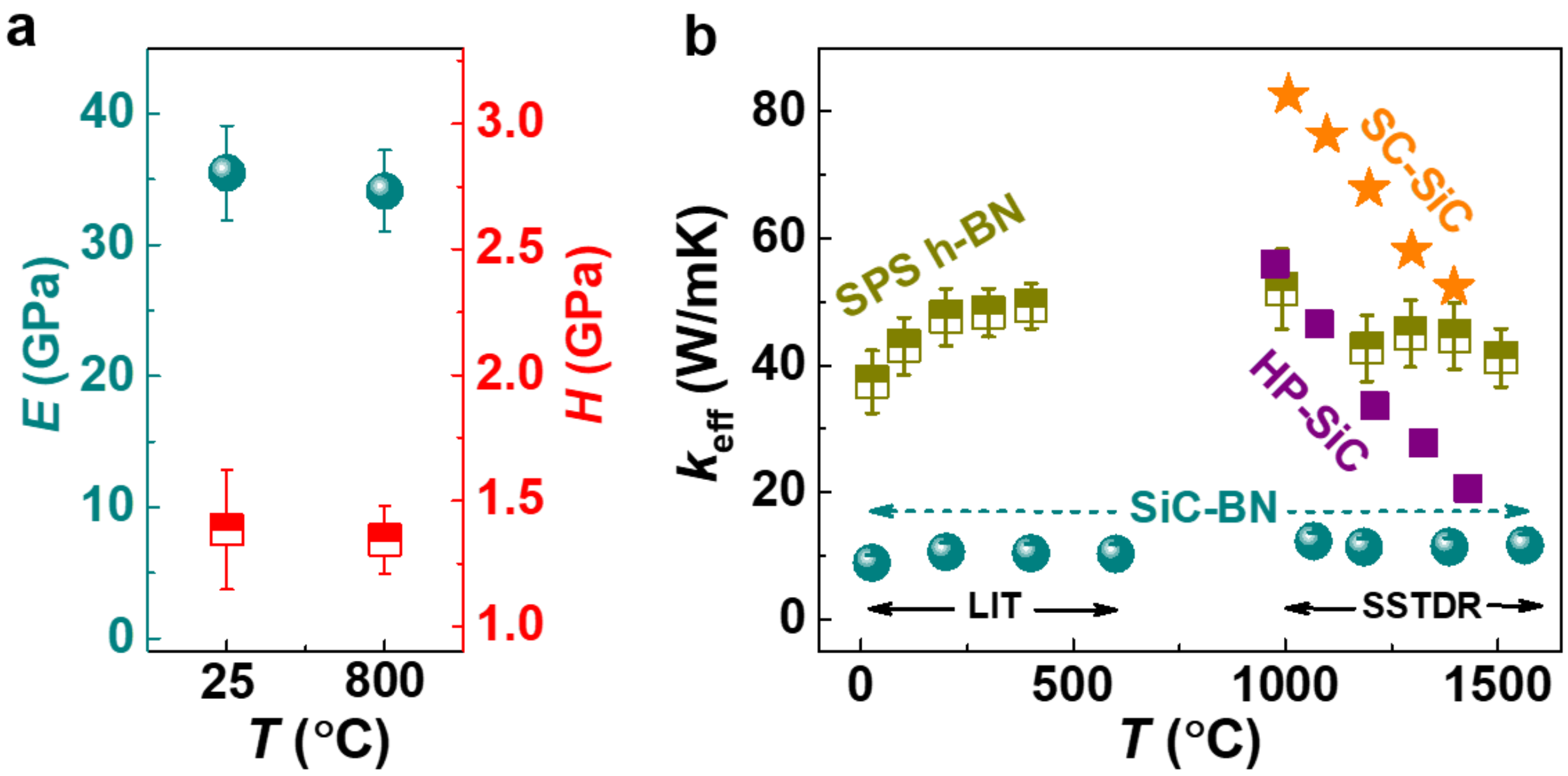

**Fig. 2: Temperature-dependent mechanical properties and thermal conductivity. a,** Young's modulus (*E*) and Hardness (*H*) exhibit negligible changes at 800 °C. **b,** Effective thermal conductivity ($k_{eff}$) exhibits temperature-independent behavior. From 25 to 600 °C, it was measured by using LIT (lock-in infrared thermography), and remains essentially constant. The measured $k_{eff}$ by using SSTDR (steady-state temperature differential radiometry) from ~1000 to 1560 °C, also shows temperature-independent behavior within experimental uncertainty, demonstrating stable thermal transport across this high-temperature regime. For comparison, the thermal conductivities of spark plasma sintered h-BN (SPS h-BN), single crystal SiC (SC-SiC), and hot-pressed SiC (HP-SiC) are shown. SC-SiC and HP-SiC data are adapted from S. N. Perevislov *et al*[30].

We measured the effective thermal conductivity ($k_{eff}$) over two temperature regimes by using lock-in infrared thermography (LIT)[31] and steady-state temperature differential radiometry (SSTDR)[32] (see **methods**). From 25 to 600 °C, the LIT measurements yield $k_{eff}$ of ~8.9-10.5 W/mK[24]. The central values increase modestly at lower temperatures and then remain nearly unchanged through 600 °C, within the experimental uncertainty. When the system is heated to high temperatures, remarkably, this temperature-independent nature still persists (**Fig. 2b**). The SSTDR measurements from ~1000 to 1600 °C yield $k_{eff}$ of ~11-12 W/mK. The values decrease slightly from 12.3 W/mK at ~1060 °C to 11.4 W/mK at ~1180 °C and then remain almost unchanged through ~1560 °C, indicating nearly temperature-independent behavior in the SSTDR regime within the experimental uncertainty. Remarkably, despite the ~1500 °C increase in temperature, this temperature-independent $k_{eff}$ demonstrates stable macroscopic heat transport under extreme thermal conditions.

The temperature-independent $k_{eff}$ is particularly interesting as it differs from the strong decrease expected when heat transport is governed predominantly by intrinsic phonon-phonon scattering[33]. The composite contains a complex network of SiC and h-BN nanograins, grain boundaries, interfaces, and pores distribution. This introduces additional scattering, interfacial resistance, and constrictions in the heat-flow pathways (see additional discussions in the **Supplementary note**). Consequently, heat transport is strongly limited by phonon scattering at structural defects and boundaries, together with interfacial and constriction resistances imposed by the composite architecture, reducing the influence that a temperature-dependent increase in intrinsic anharmonic scattering has on the macroscopic thermal conductivity[33]. In other words, the microstructure does not simply reduce $k_{eff}$, but it can also suppress much of the temperature-dependence that would otherwise originate from the constituent lattices.

## High-velocity oxygen-fuel (HVOF) torch impact test

To evaluate composite's behavior under extreme kinetic and thermal conditions, the flame produced by an high-velocity oxygen-fuel (HVOF) torch (containing oxygen and kerosene)[34] is focused on the disk surface where the hot gasses impact at speeds of ≥2000 m/s, hypersonic Mach 6)[35] (see **methods** and **Fig. 3a**). HVOF test is crucial where ceramics must perform reliably under high thermal stress while maintaining structural integrity[34,35]. The 6-minute (spacecraft re-entry time) continuous impact test performed on the disk (diameter 40 mm, thickness 2.5 mm) shows that the surface peak temperature reaches ~1500 °C at the torch-impingement zone and ~1000°C around the rim (**Fig. 3b** and Supplementary **video 1**). A steep rise in surface temperature (within 10 s) after being exposed to the torch signifies that the composite absorbed heat rapidly. Surprisingly, at the same HVOF conditions, commercial or SPS 6H-SiC undergoes catastrophic failure within 3-4 s of exposure (**Fig. 3c** and Supplementary **video 2** and **3**), whereas SPS h-BN disk starts eroding, making a crater after a few minutes (**Fig. 3c** and Supplementary **video 4**). In contrast, the composite shows its remarkable thermal stability and structural integrity upon 6-min impact without any mass loss (mass of the disk: before test = 5.1848 g, and after test = 5.1850 g) or any visible microcrack formation (**Fig. 3c**). However, the surface turns black with the formation of a whiter layer around the torch-impingement zone. We also performed the thermal cycling test (3-cycles with torch-in duration of 6 min in the heating part and 10 s of torch-off time between each cycle) (**Fig. 3d** and Supplementary **video 5**). The composite shows excellent thermal-shock resistance during thermal cycling (temperature dropping from 1500 to 300 °C in 10 s) without any crack formation, however, creating similar whiter regions and a slight increase in the mass (before test = 5.3263 g; after test = 5.3755 g) (**Fig. 3e**). The shallow modification of the surface material makes it possible to mechanically polish out the impact surface and ready to be reused (Supplementary **Fig. 12**). The composite was subjected to the same impact test targeting the surface temperature to reach ~2000 °C (by reducing the torch-to-disk distance), however, after 20 s it shattered (Supplementary **Fig. 13** and **video 6**).

We employed finite element analysis (FEA) to investigate the temperature distribution throughout the disk by performing a steady-state heat-conduction FEA on a mesh that is geometrically faithful to the imaged phase distribution (see **methods**). We obtained the steady-state front-face surface temperature of ~1504 °C at the center (with a heat-flux of ~1.21 $MW/m^2$) (**Fig. 3f** and Supplementary **Fig. 14**). The hot-zone is clearly localized at the impingement point, with a smooth radial decay that follows the shape of the input Gaussian flux distribution. The iso-contours are concentric circles aligned with the axis of symmetry, as expected from the axisymmetric flux and boundary conditions, and provide direct visual confirmation. The temperature at the rim is ~998 °C, close to the experimental one. The back face never sees the flame; and heat reaches it only by through-thickness conduction revealing the temperature distributions through the bulk of the disk (Supplementary **Fig. 15**), with the back-face center temperature of ~1385 °C.

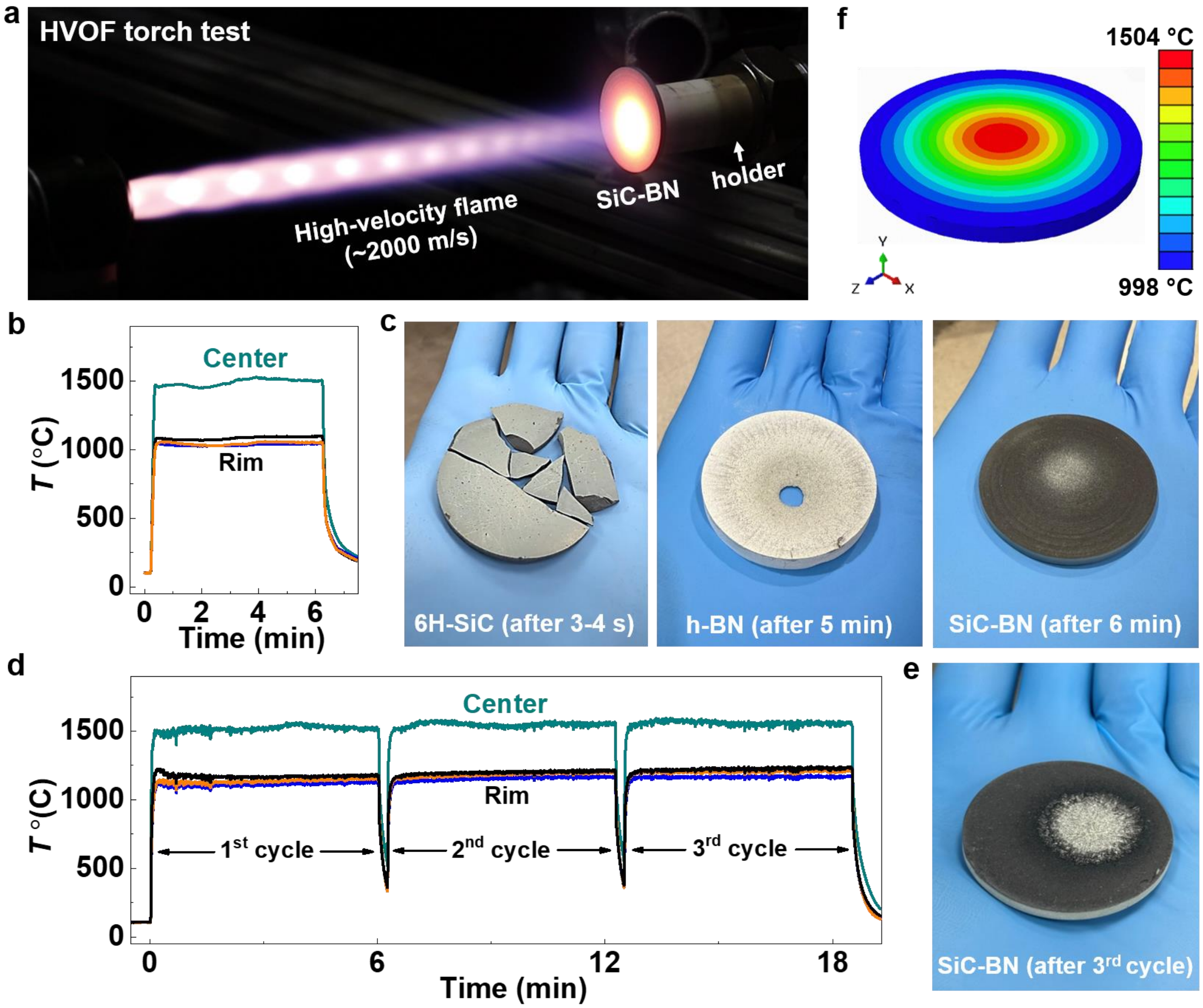

**Fig. 3: High-temperature high-velocity torch impact test. a,** High-temperature, high-velocity (~2000 m/s) impact tests were conducted in air using a high-velocity oxygen-fuel (HVOF) torch. The disks were securely held by aluminum tubes through vacuum suction. **b,** Measured surface peak temperature of ~1500 °C at the center of the SiC-BN disk, and ~1000°C near the rim. **c,** Impact tests performed at the same HVOF conditions show that monolith 6H-SiC shatters within 3-4 s of exposure, while h-BN starts eroding within a minute forming a crater. In contrast, SiC-BN composite remains stable even after 6 min of continuous torch exposure, forming a white surface layer (~few mm area). **d, e,** Thermal-cycling tests (6 min heating followed by a 10 s torch-off interval between each cycle) demonstrate excellent thermal-shock resistance of the composite. **f,** Finite element analysis steady-state front-face temperature distribution shows that the temperature reaches ~1504 °C at the center, and ~998 °C towards the rim, respectively.

## Post-characterizations of the composite after HVOF test

We characterized the top-surface white layer formed due to a 6-min impact. XPS survey scan and time-of-flight secondary ion mass spectrometry (TOF-SIMS) of that white layer show the major presence of Si and O, with minute signals from B, C, or N, indicating it is a Si-O-rich layer (Supplementary **Figs. 16** and **17**). Boron appears few nm beneath the surface as B-O species are more volatile than Si-O at higher temperatures[36,37]. To obtain the quantitative compositional information we performed EPMA. The torch impact surface has two regions; the center white layer and the black layer surrounding it. The EPMA back scattered electron (BSE) image of that white layer shows glass-like topography (**Fig. 4a**), and the corresponding compositional map (**Fig. 4b,c** and Supplementary **Fig. 18**) shows presence of elements with an obtained wt% of 28.03% (Si), 48.38 % (O), 15.18% (N), 6.81% (C), and 0.65% (B) (Supplementary **Table 3** and **4**), confirms the formation of $SiO_2$-rich top layer with some presence of C and N. The BSE image and compositional map from the black layer (Supplementary **Fig. 19**) confirms the presence of elements with an obtained wt% of 25.92% (Si), 48.66 % (O), 15.25% (N), 9.10% (C), and 0.47% (B), indicating that heat is distributed laterally throughout the surface. Cross-sectional imaging reveals that the depth of the white layer is ~150 µm (Supplementary **Fig. 20**). The cross-sectional map of the white layer (Supplementary **Fig. 21**) shows the presence of elements with an obtained wt% of 41.08% (Si), 20.21% (O), 24.20% (N), 3.50% (C), and 11.24% (B), indicating that the $SiO_2$-rich layer is only formed at the top surface. Compositional map below the 150 µm white layer shows the presence of elements with an obtained wt% of 29.77% (Si), 2.45 % (O), 30.87% (N), 24.37% (C), and 13.83% (B), similar to the pristine composite (Supplementary **Table 4**). Cross-sectional FESEM also confirms the retention of pristine-like morphology beneath the white layer (Supplementary **Fig. 22**).

To confirm the crystallinity of the white layer, the grazing incident angle XRD (GIXRD) was employed that display a broad hump, characteristic of amorphous silica (a-$SiO_2$) with a peak at ~21.9° (**Fig. 4d**). This peak corresponds to $\alpha$-cristobalite $SiO_2$ (101) (JCPDS: 01-076-0938), a crystalline polymorph of $SiO_2$ (also observed for the thermal-cycling test, Supplementary **Fig. 23**)[36]. Quantitatively, the $\alpha$-$SiO_2$ phase crystallinity is ~16% (inset of **Fig. 4d**). Bulk XRD, however, shows retention of both the pristine phases (Supplementary

**Fig. 24**), again confirming the formation of a surface silica layer, while the bulk remains the same. To corroborate this, we performed in-air furnace oxidation, and as seen, with annealing at ~1600 °C, high amounts of Si and O are found on the top surface[23], however, without any volume shrinkage or visible cracking (Supplementary **Fig. 25**).

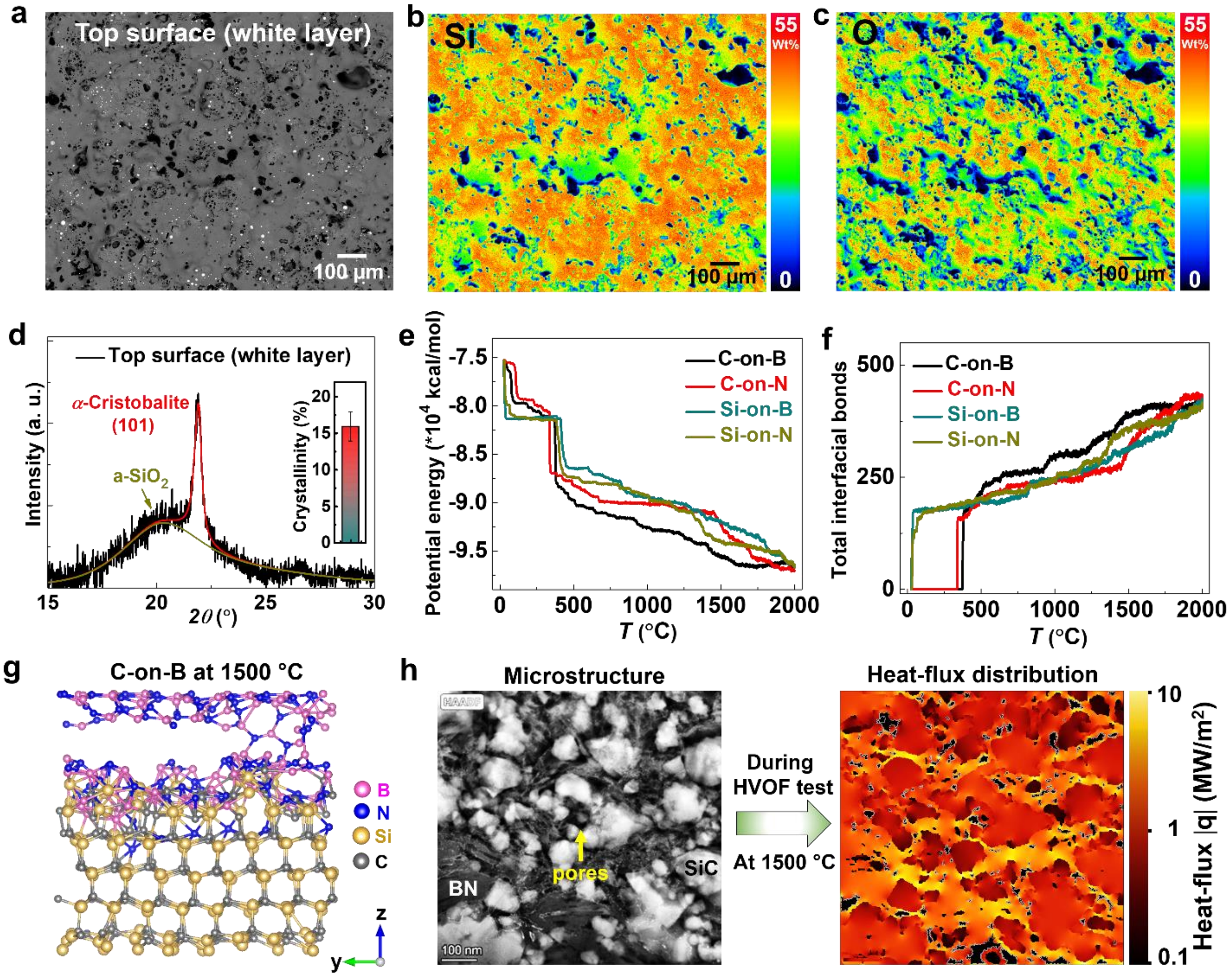

**Fig. 4: Composite surface after the HVOF impact, atomic-scale interfacial stability, and finite element analysis thermal-transport. a,** EPMA backscattered electron image of the top white layer reveals glass-like topography. **b, c,** WDS elemental maps show the enrichment of Si and O within the surface layer. **d,** Grazing incident angle x-ray diffraction (GIXRD) from the white top surface shows a diffraction peak at ~21.9 °C, corresponds to $\alpha$-cristobalite $SiO_2$ (101), around an amorphous silica (a-$SiO_2$) background, with $\alpha$-$SiO_2$ crystallinity of ~16%. **e-g,** ReaxFF molecular dynamics simulations of thermally-induced reconstruction at C-on-B, C-on-N, Si-on-B, and Si-on-N interface configurations up to

2000 °C. **e,** Evolution of the system potential energy with temperature, **f,** corresponding changes in the total number of interfacial bonds. Sharp transitions indicate the onset of interfacial reconstruction, showing C-on-B exhibits higher resistance to thermally induced interfacial reconstruction. **g,** Side-view atomic configuration of the most energetically favorable reconstructed C-on-B interface at 1500 °C **h,** Thermal finite-element analysis (FEA) heat-flux map at 1500 °C reconstructed from the HAADF microstructure (gray: SiC, black: BN), showing pores (darker regions) interrupting the local heat flow, thereby reducing the thermal stress and contributing to the thermal-shock resistance.

## Atomistic insights into high-temperature interfacial stability

To obtain atomistic insights about the interfacial bonding's network at 1500 °C (i.e. the HVOF condition), we investigated temperature-dependent evolution of the interfaces via ReaxFF by continuously heating the systems up to 2000 °C, while monitoring the potential energy and evolution of interfacial bonding. The onset of new interfacial bond formation was used as a measure of the resistance of the initially weak van der Waals (vdW)-bound interfaces to thermally induced reconstruction. The Si-on-B and Si-on-N interfaces show early bond formation at ~33° and ~38 °C, respectively, whereas C-on-N and C-on-B retains their initial interfacial character to much higher temperatures, with reconstruction initiating at ~341° and ~382 °C, respectively (**Fig. 4e,f**). Therefore, C-on-B interface exhibits the highest reconstruction-onset temperature, indicating the higher resistance to thermally-induced interfacial reconstruction.

Following the onset of bonding, all interfaces progressively developed increasing numbers of interfacial bonds, reflecting continued chemical integration of the interfaces with increasing temperature. The potential-energy evolution provides a complementary measure of the relative energetic stability of these reconstructed interfaces. At 1500 °C, the C-on-B configuration exhibits the lowest potential energy among the four interfaces, indicating that it forms the most energetically favorable reconstructed state under these conditions. Deconstructing these energetic minimums reveals that the enhanced stability is associated with the simultaneous formation of C-B and Si-N bonds. Specifically, at 1500 °C, the C-on-B configuration exhibits a high number of C-B bonds while simultaneously developing a substantial population of Si-N bonds, comparable to or higher than those observed in the other configurations (Supplementary **Fig. 26a-d**). The four interface configurations at 1500 °C are shown (**Fig. 4g** and Supplementary **Fig. 26e-g**), exhibiting interfacial reconstructions. The concurrent formation of chemically favorable C-B and Si-N interfacial bonds provides a molecular-level explanation for the lower potential energy of the C-on-B interface, contributing to stabilize the reconstructed interface, and forming a Si-N/C-B-rich chemical bonding network that helps to mitigate structural degradation.

## Discussion on extreme condition thermal stability

The composite exhibits remarkable thermal stability under high-temperature, high-velocity flame exposure, maintaining its structural integrity while monolithic 6H-SiC shatters and h-BN undergoes erosion. The shattering of 6H-SiC is primarily driven by its high mechano-thermal stresses due to low fracture toughness ($K_c$) (Supplementary **Table 5**)[38]. The temperature gradients imposed by impact induce high thermal-shock, making it vulnerable for crack propagation almost instantaneously, resulting in catastrophic failure. For h-BN, the erosion is largely due to the combined effects of its layered structure and easy oxidation. Its relatively weak interlayer bonding makes it susceptible to shear and material removal under high-velocity flow. h-BN oxidizes in air at ≥800 °C, forming a low temperature melting point glass and volatile $B_2O_3$ species[36,39]. This continuous formation and removal of $B_2O_3$ chemical species accelerates mass loss. Hence, instead of fracturing catastrophically like brittle 6H-SiC, h-BN undergoes erosion over time.

In contrast, the superior thermal stability of the composite arises from a synergistic interaction between the two phases that mitigate the weaknesses each material exhibits individually. The 6H-SiC provides high mechanical strength, whereas the h-BN phase due to its layered nature, contributes to thermal-shock resistance by accommodating strain and reducing the buildup stresses during rapid heating. Under oxidizing conditions, they produce a viscous, glassy thermal-shock resistant silica layer (containing amorphous and $\alpha$-cristobalite phase) against further oxygen diffusion and shields the material from further oxidation. In addition, porosity plays a key role in enhancing thermal stability as pores slow down the heat transfer into the bulk and minimizes temperature differentials that can lead to crack formation. XRM shows that pores are still present in the bulk even after the impact (Supplementary **Fig. 27**). By using the high-angle annular dark-field scanning transmission electron microscopy (HAADF-STEM) image as the geometric input we mapped the heat-flux distribution within the microstructure at 1500 °C (see **methods**). The heat-flux map shows that pores act as flux obstacles (**Fig. 4h**), deflecting streamlines into the "tortuous" pattern typical of porous ceramics, which reduces thermal stress[40]. A low coefficient of thermal expansion (CTE) of ~$2.9\times10^{-6}$ /°C (measured via push-rod dilatometry) of the composite further limits thermally-induced strain. Hence, together with surface silica layer, porous microstructure, and a low CTE favors in maintaining an effective heat-transport network, enhancing the thermal-shock resistance.

In summary, we fabricated lightweight SiC-BN ceramic composites exhibiting excellent thermal-shock resistance at extreme environments. In addition, it shows temperature-independent thermal conductivity up to ultrahigh-temperature, providing an additional thermo-physical characteristic relevant to retaining structural integrity and maintaining a predictable bulk thermal response under extreme thermal loading. Unlike monolithic high-temperature ceramics having high-density, brittleness, and lack of machinability (limiting their practical use in complex and/or weight-sensitive applications), this easily machinable composite's lightweight, stable thermal-transport and thermal-shock resistance, makes it useful in aerospace propulsion, combustion engines, and advanced manufacturing related high-temperature structural components, where weight and design flexibility with structural integrity are more critical than absolute ultra-high temperature capability.

## Methods

### Spark plasma sintering of SiC-BN composites

The SiC-BN composites were prepared via spark plasma sintering (SPS) by using commercially available 6H-SiC (particle sizes ~100 nm, Purity: 98.5%, Batch Number. 10023B3 MSE suppliers, Tucson, AZ, USA) and h-BN powders (particle sizes ~800 nm, Purity: 99.8%, CAS#: 10043-11-5, US Research Nanomaterials, Inc., Houston, TX, USA). SPS was conducted using an SPS 25-10 system (Thermal Technology LLC, California, USA) at Texas A&M University, USA. For all the sintered materials, a constant uniaxial pressure of 90 MPa was applied, with a heating rate of 50 °C/min, reaching a maximum temperature of 1800 °C. For SPS, equal wt% of 6H-SiC and h-BN powders were mixed and ground for 20 min in a mortar pestle. The powder was wrapped in graphite foil and kept inside the graphite die and placed into the SPS chamber under an initial pressure of 5 MPa. Graphite foil serves as a barrier to prevent reactions between the materials and die, while also enhancing thermal distribution and insulation. The chamber was pre-evacuated to ~$2\times10^{-5}$ Torr for 30 min, followed by sintering for 1 hr at 1800 °C under ultra-high purity argon gas (5N, ~99.999%) atmosphere. During SPS, a direct current passes through the sample, generating the Joule heating process. The temperature during the sintering process was monitored using an optical pyrometer (Raytek, Berlin, Germany, model D-13127). After sintering, the pressure was released gradually at ~5 MPa/min, while the temperature was ramped down at ~100 °C/min. After sintering, we polished the disks to remove the graphite foil and mechanically cut the samples into desired sizes for various measurements. For comparison, we also fabricated pure 6H-SiC and h-BN disks by using the same precursor powders and SPS conditions. The density was measured by using the solid cylinder method.

### Structural characterizations (XRD, TGA, XPS, Raman spectroscopy)

X-ray diffraction (XRD) were carried out using a Rigaku SmartLab thin-film diffractometer (Tokyo, Japan), operating at 40 kV and 40 mA with a monochromatic Cu $K_\alpha$ radiation source (λ = 1.5406 Å), and a scan rate of 1°/min (for bulk scan) and 0.2°/min (for grazing angle). For the grazing angle XRD, the incident beam angle ($\theta$) was kept at 0.1°. For the in-plane XRD, we used the same XRD instrument and performed 2-Theta Chi/Phi scans. High-temperature XRD in air was carried out in Rigaku SmartLab II x-ray diffractometer from 25 to 1000 °C, with 100 °C interval and dwelling time of 20 min. The scan rate was 5°/min. The sample was cut into the desired shape/size to fit into the commercially available Pt holder (SH Tray, Pt, High Temp, for AL203 SH, Rigaku Americas, TX, USA). We carried out X-ray photoelectron spectroscopy (XPS) using a PHI Quantera SXM scanning X-ray microprobe with a monochromatic Al $K_\alpha$ source (1486.6 eV). High-resolution core-level spectra were collected at a pass energy of 26 eV, and spectral fitting was performed with MultiPak V6.1A software. Raman measurements were obtained using a Renishaw inVia confocal microscope equipped with a 532 nm excitation laser. Thermogravimetric analysis (TGA) oxidation-resistance was done in air with the Mettler Toledo TGA/DSC 3+ system up to 1000 °C with a ramp up rate of 10 °C/min.

**X-ray microscopy (XRM)**

X-ray microscopy (XRM) images with a voxel size of ~1.5 µm were collected using a Sigray Eclipse 900 XRM system, which features an ultrahigh-resolution nanofocus X-ray source operating at 60 kV and a 27-MP large field-of-view flat-panel detector. The acquired data were then processed using 3D Dragonfly software, with a cropped region shown for improved visualization. Phase segmentation to distinguish the pores were carried out using the machine learning (ML) algorithm integrated within the software.

**Time-of-flight secondary ion mass spectrometry (TOF-SIMS)**

Positive high mass resolution depth profile was performed using a TOF-SIMS NCS instrument, which combines a TOF.SIMS 5 instrument (ION-TOF GmbH, Münster, Germany) and an in-situ Scanning Probe Microscope (NanoScan, Switzerland) at Shared Equipment Authority from Rice University.

The analysis field of view was 150×150 $\mu m^2$ ($Bi_3^+$ @ 30keV, 0.3pA) with a raster of 128 by 128 along the depth profile. A charge compensation with an electron flood gun has been applied during the analysis. An adjustment of the charge effects has been operated using a surface potential. The cycle times were fixed to 100 µs (corresponding to m/z = 0

– 911 a.m.u mass range). The sputtering raster was 450×450 µm$^2$ ($Cs^+$ @ 1keV, 85 nA). The beams were operated in non-interlaced mode, alternating 2 analysis cycles and 0.5 per frame of sputtering followed by a pause of 5s for the charge compensation.

The $MCs^{n+}$ ($n$ =1, 2) depth profiling has been also used for improving the understanding of the data. This is a useful method, mainly applied to quantify the alloys but also to identify any ion compounds. The cesium primary beam is used for sputtering during the depth profile and permits to detect $MCs^+$ or $MCs^{2+}$ cluster ions where M is the element of interest combined with one or two Cs atoms. The advantages of following $MCs^+$ and $MCs^{2+}$ ions during ToF-SIMS analysis include the reduction of matrix effects and the possibility of detecting the compounds from both electronegative and electropositive elements and compounds.

All depth profiles have been point-to-point normalized by the total ion intensity and the data have been plotted using a 10-points adjacent averaging. Both normalization and smoothing have permitted a better comparison of the data from the different samples. The depth calibrations have been established using the interface tool in SurfaceLab version 7.3 software from ION-TOF GmbH to identify the different interfaces and based on the estimated thickness of the BN layer.

**Electron microscopy (FESEM and HRTEM)**

Cross-sectional sample preparation and characterizations were performed using focused ion beam-scanning electron microscopy (FIB-SEM) on a FEI Helios 660 Dual Beam FIB-SEM system equipped with an energy-dispersive X-ray spectroscopy (EDS) detector for elemental analysis. The cross-sections were milled using a 30 keV $Ga^+$ ion beam for initial trenching, followed by cleaning cross-section milling to improve surface quality. SEM imaging was conducted at an accelerating voltage of 5 keV with a probe current of 0.1 nA at a stage tilt of 52°. Vertical streaks observed in the SEM images are attributed to curtaining artifacts associated with ion-beam milling.

For high-resolution transmission electron microscopy (HRTEM), lamellae were prepared from selected region using the same FEI Helios 660 Dual Beam FIB-SEM system equipped with a $Ga^+$ ion source. Initial coarse milling was performed at 30 keV, followed by progressive thinning with decreasing accelerating voltages (30-5 kV) and appropriate beam currents to achieve electron transparency while minimizing ion-induced damage. The lamellae were extracted using a standard in situ lift-out procedure and subsequently mounted onto Cu TEM grids (Ted Pella). Final low-energy polishing was applied to ensure electron transparency. HAADF-STEM imaging was carried out using an aberration-corrected FEI Titan Themis[3] operated at an accelerating voltage of 300 kV.

**Electrical transport**

Physical property measurement system (PPMS) from quantum design, was used to measure the DC resistivity of a rectangular sample as a function of temperature, ranging from 2-300 K, at a rate of 1 K/min. For the resistivity, a constant current of 5 mA was applied along the in-plane and the voltage (V) was measured. We used the standard four-point probe method with silver point contacts.

**Machining of the disk**

For pattering, we used a 60W MOPA fiber laser fitted with a 275 mm lens and multiple passes to etch the material away. We used Lightburn which has built in settings for slicing a grayscale image into different layers based on the shadowing. For slicing the disks, we used a diamond blade, whereas for drilling we used a tungsten carbide drilling machine.

**Mechanical characterizations**

Nanoindentation experiments were performed at room temperature and 800 °C (up to the instrumental limit) inside a scanning electron microscope (SEM; FEI QUANTA 3D). A tungsten carbide (WC) Berkovich tip mounted on a high-temperature nanoindenter (KLA Instruments, InSEM HT) was used for all measurements. The tests were carried out at a constant indentation rate of 0.2 /s up to 40 mN, and the Oliver-Pharr method[41] was applied to determine the Young's modulus and hardness. For the high-temperature experiments, only measurements obtained at a thermal drift rate below 1 nm/s were included to report the average modulus and hardness values.

An iMicro indenter (KLA Instruments) was utilized to determine the fracture toughness of polished BN and SiCBN disks. Indentation tests were carried out at applied loads of 1000 mN and a dwell time of 5 s to produce well-defined impressions on the sample surfaces. From the optically measured crack and impression lengths, the fracture toughness was estimated using the Lawn-Evans-Marshall (LEM) model[42]:

$$K_c = A(\frac{E}{H})^{2/3}\frac{P}{c^{3/2}}(\frac{a}{l})^{1/2},$$

where *A* is an empirical constant taken as 0.016 for a sharp Berkovich tip, *E* is the Young's modulus, *H* is the hardness, and *P* is the applied load. In addition, *c* represents the total crack length, while *a* represents the indentation size, and *l* denotes the crack length measured from the indentation corner to the crack tip (i.e., $c = a + l$).

Micropillar compression tests were carried out to evaluate the mechanical behavior of SiC-BN disk at room temperature. The micropillars were fabricated using focused ion beam (FIB) milling inside a scanning electron microscope (SEM; Quanta 3D FEG, FEI). The diameters of SiC-BN micropillars ranged from 3.0-4.4 μm with taper angles no larger than 2.3° to ensure homogeneous deformation. The dimensions of the micropillars are summarized (Supplementary **Table 2**). The compression tests were carried out at a constant strain rate ($10^{-3}$ /s) using a nanoindenter (InSEM, KLA) equipped with a 20 μm-diameter diamond flat punch tip.

### Reciprocal wear test

The Rtec Tribometer (Model: MFT-5000, California, USA) was used to perform the wear test. The indenter tool was Yttria-Stabilized Zirconia (6 mm diameter) ball. Reciprocal tests were carried out under fixed load conditions (1N and 5N) up to 2 mm, with a velocity of 4 mm/min. Topographic image analysis was performed after the scratch to measure the areal profile and indentation depth, and to calculate the wear rate and wear volume.

**Dilatometry tests**

Dilatometry test was done in a DIL 402 C dilatometer (Netzsch) using an alumina push rod and 15 cN of pushing load. It was performed from room temperature up to 1300 °C. The heating/cooling rate was 5 °C/min. The thermal expansion coefficient was calculated from the slope of the linear regions of the relative dilation curves ($(L-L_0)/L_0$).

**Temperature-dependent thermal conductivity**

The effective thermal conductivity $k_{eff}$ of the SiC-BN composite was measured over two temperature regimes using lock-in infrared thermography (LIT) from room temperature to 600 °C and steady-state temperature differential radiometry (SSTDR) from ~1000 to ~1600 °C. The sample used was ~1 mm thick. Although the SiC particles, h-BN flakes, interfaces, and pores are locally heterogeneous, their random distribution at the macroscopic scale allows the bulk composite to be treated as an effectively isotropic medium for the thermal measurements. We also measured the thermal conductivity of SPS h-BN solids by using the similar method.

**Lock-in infrared thermography.** LIT measurements were performed using a 532 nm laser to periodically heat the sample surface (Supplementary **Fig. 28**). The laser intensity was modulated with a mechanical chopper at frequencies between 1.5 and 2.0 Hz, with a laser spot diameter of 840 $\pm$ 20 µm and incident powers between 0.3 and 0.6 W. The resulting periodic surface-temperature response was measured using an Infratec ImageIR 8380 infrared camera operating over a spectral range of approximately 2-5 µm. The camera was synchronized with the laser modulation frequency to obtain the phase-resolved thermal response while suppressing the steady heating background, signal drift, and camera noise. Sample temperature was controlled using an Instec HCP421V heating stage. Argon was continuously flowed through the measurement chamber to minimize oxidation, with the regulator at the gas tank set to 2 psi. A $CaF_2$ window provided optical access within the infrared-camera spectral range.

The temperature response was analyzed using the transient heat equation in cylindrical coordinates. Treating the composite as macroscopically isotropic,

$$C\frac{\partial T}{\partial t} = k_{eff}\left(\frac{\partial^2 T}{\partial r^2} + \frac{1}{r}\frac{\partial T}{\partial r} + \frac{\partial^2 T}{\partial z^2}\right)$$

where $C$ is the volumetric heat capacity. Temperature-dependent heat capacities were obtained from literature values for SiC and h-BN[43], combined according to the 50-50 wt% composite composition and converted to volumetric heat capacity using the measured composite density. For harmonic laser heating, the heat equation was transformed in the radial direction using a zeroth-order Hankel transform and solved using a multilayer transfer-matrix formalism. The calculated radial phase response was fitted to the measured phase delay as a function of distance from the heating spot to determine $k_{eff}$.

**Steady-state temperature differential radiometry.** SSTDR measurements were performed in the range of ~1000-1600 °C under 1 atm argon, with the oxygen pump operating during the measurements to suppress oxidation. A continuous-wave 1030 nm laser (Trumpf TruDisk 4001) with a 5.8 mm diameter spot was used to heat the sample to the desired background temperature (Supplementary **Fig. 29**). A second continuous-wave 532 nm laser (Coherent Verdi V5), focused to an approximately 850 µm diameter spot at the center of the heated region, induced a small additional temperature rise, $\Delta T$. The smaller Gaussian perturbation beam was positioned within the approximately uniform-intensity region of the larger heating spot so that the measured temperature rise was not convoluted with the spatial variation of the background heating profile. The absorptance at the 532 nm perturbation-laser wavelength was also measured directly using a visible pyrometer (FAR FMP2/2x) to find the absorbed power that causes the measured $\Delta T$.

The measured differential temperature field was analyzed using the axisymmetric two-dimensional steady-state heat equation,

$$k_{eff}\left(\frac{\delta^2 T}{\delta r^2} + \frac{1}{r}\frac{\delta T}{\delta r} + \frac{\delta^2 T}{\delta z^2}\right) = -Q(r, z)$$

with the absorbed laser fluxes ($Q$), laser spot dimensions, sample geometry, and radiative losses incorporated through the boundary conditions. This heat equation was solved numerically with nonlinear radiative losses where heat flux depends on both temperature and its gradient, and $k_{eff}$ was determined by reproducing the experimentally measured laser-induced temperature rise.

### High-velocity oxygen fuel (HVOF) torch impact test

HVOF tests were performed on the composite disks with diameter 40 mm and thickness 2.5 mm. The sample was held at the end of a 25 mm diameter alumina tube using vacuum pressure and was exposed to high temperature combustion gases from the nozzle of a commercial High Velocity Oxygen-Fuel torch (JP 5220, Praxair TAFA). This torch uses premixed oxygen and kerosene with combustion gases accelerated to supersonic speeds through a de Laval type nozzle. The oxygen and kerosene flows were 1925 SCFH (908.5 slpm) and 7.2 GPH (0.45 lpm) respectively, producing an estimated gas velocity of ~Mach 6 (~2000 m/s) during impact. The distance between the torch and the disk for the 6 min test and for the thermal cycle was 20 cm, with the torch position controlled using an industrial six-axis robot. Two SiC-BN samples were subjected to 1 and 3 cycles respectively with a torch-in duration of 6 min in the heating part of the cycle and 10 s of torch-off time between each cycle. The temperature on different surface locations was monitored using a long wave infrared (IR) thermal imaging camera (A655SC, FLIR Systems, Wilsonville, OR, USA). The emissivity was set as 0.89. A constant flow of air cooling was provided on the disk back surface to cool down the samples. We used the similar parameters for the pure 6H-SiC and h-BN disk impact tests.

### Electron probe microanalysis (EPMA)

Quantitative chemical analyses were performed using a JEOL JXA-8530F Hyperprobe equipped with a field-emission-assisted thermionic (Schottky) emitter, five wavelength-dispersive spectrometers (WDS), and one silicon-drift energy-dispersive spectrometer (SD-EDS). All quantitative analyses and WDS maps were acquired at 10 kV accelerating voltage and 50 nA beam current. Both wavelength-dispersive spectrometry (WDS) point analyses and WDS X-ray compositional mapping were carried out. Quantitative data reduction was performed using the CITZAF matrix correction software (Armstrong, 1995), implemented in the JEOL on-line environment.

Point analyses were acquired for the elements B, C, N, O, and Si, all measured as $K_\alpha$ X-ray lines. C, B, N, and O were measured on layered dispersion element (LDE) pseudo-crystals, appropriate for light element detection: LDE2 for B and C, and LDE1 for N and O. Si was measured on a thallium acid phthalate (TAP) crystal. Peak counting times were 10 s for all elements, with background counting times of 5 s on each side of the peak. The following references were used for calibration: C: graphite (C_graphite_SPI; 99.99 wt% C), B: boron nitride (BN_SPI; 43.6 wt% B), N: boron nitride (BN_SPI; 56.44 wt% N), O: quartz (Quartz_SPI; 53.26 wt% O), and Si: quartz (Quartz_SPI; 46.74 wt% Si).

WDS compositional maps were acquired for B, C, N, O, and Si ($K_\alpha$ lines) by stage scanning. The maps are displayed as concentration (wt%) color images using a thermal color scale. The acquisition conditions, read from the instrument metadata, are as follows: Accelerating voltage: 10.0 kV, beam current: ~50 nA, spot (focused beam, ~ 80 nm in diameter), Dwell time 200 ms. The size of the maps is 650 × 500 pixels, with step size 0.60 µm (X) × 0.60 µm (Y). Total mapped area: 390 µm (X) × 300 µm (Y), unidirectional stage scanning. Details on spectrometer and detector setting are presented (Supplementary **Table 6** and **7**).

Following EPMA analysis of the white surface layer and the surrounding dark area of the disk, a cross-sectional specimen was prepared for further EPMA characterization. Sample preparation consisted of two sequential steps: (a) sectioning the disk along its diameter using a precision diamond saw, and (b) mounting the resulting cross-section in epoxy resin (EpoxyCure EpicureTM 2, ref. 20-3430-064, Buehler). After a curing period of 24 hours, the mounted specimen was subjected to a multi-stage mechanical preparation procedure comprising three grinding stages using silicon carbide (SiC) abrasive, followed by three successive polishing stages: an initial polish with 1200 grit SiC paper, an intermediate polish using alumina suspension with a mean particle size of 1 µm, and a final polish using alumina suspension with a mean particle size of 0.3 µm.

Prior to EPMA analysis, all specimens were coated with a 25 nm-thick amorphous carbon film to ensure adequate electrical conductivity.

**Reactive force-field (ReaxFF) molecular dynamics (MD) simulations**

To investigate the atomic-scale stability of the 6H-SiC/h-BN interface, ReaxFF molecular dynamics simulations were performed using a force field constructed by merging previously developed Si/C/N and C/B/N parameter sets[44]. The Si-B interaction parameters were further trained against the heat of formation of the stable $Si_3B$ phase and its equation-of-state data, including the bulk modulus and elastic constants data[45]. A successive single-parameter search optimization was performed to optimize the parameters in ReaxFF and minimize the error using the following equation[46]:

$$Error = \sum_{i=1}^{n} \left[\frac{x_{i,QM} - x_{i,ReaxFF}}{\sigma}\right]^2$$

where σ denotes the weight assigned to every data point according to its relative significance in the optimization process, *n* is the total number of data points in the training set, and *x* represents an individual data point in the training set, specifically the energy in this study.

The heat of formation of stoichiometric $SiB_3$ was included in the training set, showing good agreement between ReaxFF and quantum mechanics (QM) data (Supplementary **Fig. 30a**). The equations of state (EOS) for $SiB_3$ were also used in training by systematically expanding and compressing the lattice volume around the ground-state value to capture both compressed and thermally expanded states. In addition, uniaxial strain energies were evaluated. ReaxFF reproduces the QM results reasonably well near equilibrium and captures the overall strain-dependent trends (Supplementary **Fig. 30b-e**), although larger deviations are observed at higher strains, particularly for the $C_{22}$ case.

The resulting force field accurately reproduces key condensed-phase properties across Si-C, Si-N, Si-B, C-B, and C-N systems, including heats of formation and the relative stability of different phases. However, the SiC/BN interfacial energies considered in this study were not explicitly included in the ReaxFF training set. Therefore, while the calculated interface energies provide a useful assessment of the relative stability of the investigated interface configurations, their quantitative accuracy should be further validated against DFT calculations in future work.

Heterostructure models of the 6H-SiC/h-BN interfaces were constructed by cleaving the polar 6H-SiC (0001) surfaces along the single-bond cleavage planes, resulting in Si- and C-terminated surfaces with one dangling bond per surface atom[47,48]. To minimize the in-plane lattice mismatch while maintaining periodicity in the x and y directions, the SiC and h-BN surface unit cells were replicated 4×4×1 and 5×5×1, respectively. Both polarities of the SiC surface were considered: the Si-terminated (0001) surface and the C-terminated $(000\bar{1})$ surface. For each surface polarity, different local interface configurations were examined by laterally translating the h-BN layer relative to the SiC surface. The configurations were defined according to the atom directly beneath the reference position (0,0) of the interface: when the surface Si (or C) atom at this reference position was aligned with an N or B atom of h-BN, the configuration was designated "on-top-of-N" or "on-top-of-B," respectively.

The resulting heterostructures were placed at the center of a periodic simulation cell with dimensions of 12.306 × 21.315 × 100 Å. A large vacuum region was introduced along the surface-normal direction to minimize interactions between periodically repeated slabs. All structures were subsequently energy-minimized to obtain their relaxed configurations and corresponding potential energies.

To evaluate the thermal stability and structural evolution of the different SiC/BN interface configurations, molecular dynamics (MD) simulations were subsequently performed using the energy-minimized structures as the initial configurations. Periodic boundary conditions were applied in the in-plane *x*- and *y*-directions, while a nonperiodic boundary condition was imposed along the surface-normal z direction to preserve the free surfaces of the heterostructure. The interface structures were placed in a simulation box with dimensions of 12.306×21.315×20 Å. An NVT ensemble was employed, and the temperature was continuously increased from room temperature to 2000 °C (2273 K) using a linear heating rate of 5 K/ps. A Berendsen thermostat with a temperature damping constant of 100 fs was used to control the system temperature during the heating process.

A timestep of 0.25 fs was used for all simulations. The temperature ramp was employed to examine the thermally induced structural evolution of the interfaces, including the formation, breaking, and rearrangement of interfacial bonds at elevated temperatures. The resulting atomic configurations and interfacial bonding behavior were analyzed to assess the relative thermal stability of the different SiC/h-BN interface configurations.

**Thermal finite element analysis**

At 1500 °C the SiC-BN composite's thermal conductivity is 11.57 W/mK, less than both constituent phases (hot-pressed SiC ~18.1, SPS h-BN ~41.8 W/mK). A two-scale Fourier finite-element analysis in ABAQUS/CAE 2024 explains this observation. A disk-scale model, inverse-calibrated to in-situ pyrometry, fixes the HVOF thermal load: an absorbed Gaussian flux of 1.21 $MW/m^2$ peak ($\sigma$ =8.9 mm) reproduces both measured temperatures to within ~5 °C. A microstructure-resolved model built from a segmented STEM image then splits the conductivity deficit. The 5 % porosity removes only 12.5%. The remaining factor of 2.12 requires a SiC/BN interfacial resistance of $R$ =2.06 × $10^{-8}$ $m^2K/W$, which pushes heat out of the SiC grains and into the continuous BN network.

**Disk-scale model, calibrating the HVOF load:** A 7,704-element DC3D10 model of the disk (diameter $r$ =20 mm, thickness $t$ =2.5 mm; measured density $\rho$ =1850 $kg/m^3$, specific heat $c_p$ =750 J/Kg-K, thermal conductivity $k_{SiC\text{-}BN}$ =11.65 W/mK, and emissivity $\varepsilon$ =0.89) was run through the 360 s heating step. The flame face receives a Gaussian flux $q''(r) = q''_{pk} \exp(-r^2/2\sigma^2)$. All faces re-radiate, the rim sits in the wall jet ($h$ =800 $W/m^2K$, $T_\infty$ =1300 °C), and the back face has free convection ($h$ =20 $W/m^2K$).

A 5 × 5 parameter sweeps over $q''_{pk}$ =0.8-3.0 $MW/m^2$ and $\sigma$ =8-20 mm makes the two-measurement inverse problem well posed. The center temperature $T_1$ tracks $q''_{pk}$ and the rim temperature $T_3$ tracks $\sigma$, so their iso-contours cross nearly at right angles (Supplementary **Fig. 14**). The calibrated point, $q''_{pk}$ =1.21 $MW/m^2$ with $\sigma$ = 8.9 mm, gives $T_1$ =1504 °C and $T_3$ =998 °C in a direct run against ~1500 and ~1000 °C measured. With $F_O = \alpha t/R^2$ ~7.6 the field is quasi-steady, so it depends on k and the boundary conditions, not on $\rho c_p$. The back-face center sits at ~1385 °C. That 120 K through-thickness drop is invisible to pyrometry and implies ~0.56 $MW/m^2$ conducted axially. Martin's impinging-jet correlation gives $h_{stap}$ ~1.56 $kW/m^2K$ and a cold-wall flux of ~4 $MW/m^2$. At a 1500 °C wall the driving temperature collapses and re-radiation removes $\varepsilon\sigma T^4$ ~0.50 $MW/m^2$. The disk absorbs ~0.55 kW, only 0.23% of the ~237 kW combustion power. One tension remains: 1.21 $MW/m^2$ at this $h_{stap}$ implies a stagnation recovery temperature near 2280 °C, above the 1700-1900 K estimated for the entrained free jet.

**Microstructure model at 1500 °C, heat-flux:** A segmented STEM image (779 × 752 nm at 2.30 nm/px) was meshed with conforming DC2D3 triangles. Its 15.8 % pore area is a projection artefact: Monte-Carlo stereology maps it to the measured ~5 vol% for a ≈67 nm foil, so pores were eroded to 5 % (50 % SiC, 45 % BN). Nodes on every SiC/BN and BN/pore boundary were duplicated per phase and recoupled by gap conductance, so each interface carries a temperature jump $q_n = (T_A - T_B)/R$. Pores hold gas (~0.108 W/mK) with a kinetic-theory wall conductance.

A uniform flux of 1.21 $MW/m^2$ enters one face, the opposite face is held at 1500 °C, and $k_{eff} = \langle q_x \rangle L_x / \Delta T$. The problem is linear, so $k_{eff}$ and R do not depend on the applied flux. With all other inputs fixed, $k_{eff}$ falls monotonically with the SiC/BN resistance, so bisection gives a unique R. Repeating it at the ±1σ bounds of the measurement gives its uncertainty.

Filling the pores raises the perfect-interface conductivity only from 24.51 to 28.01 W/mK (12.5 %). The pores are effectively insulating (treating them as vacuum shifts *R* by <1 %) and matter mainly by narrowing h-BN ligaments. Interfacial resistance of $2.06 \times 10^{-8}$ $m^2K/W$ (a conductance of ~49 $MW/m^2K$), has a Kapitza length $R \cdot k$ of ~0.86 µm in h-BN and ~0.37 µm in SiC (Supplementary **Table 8**). Crossing every interface in series would give ~1.7 W/mK; the measured 11.57 shows heat avoids most crossings. Each SiC grain costs two crossings, so SiC's share of the heat drops from 40.7% to 31.7%, while BN, at 45% of the area, carries 68.3%.

**Verification and limits:** Refining the disk mesh (7,704 to 24,643 elements) changed temperatures by <1 %. Three nested microstructure meshes (38k-342k elements) give similar results. the interface knock-down varied by <1.5 %, and the recalibrated R stayed within the ±1σ band of the measurement. the interface knock-down varied by <1.5 %, and the recalibrated R stayed within the ±1σ band of the measurement. This proves that the results are mesh independent.

- R is calibrated, so matching *k* is by construction; the evidence is R's magnitude and the flux redistribution.
- SiC conductivity is extrapolated that gives 15.9-20.0 W/mK corresponding to which R might vary from -9 % to +7 %, well inside the measurement band. SiC grain-size suppression is not modelled, so R is an upper bound.
- Transport is treated as diffusive, so sub-continuum boundary scattering is folded into R. The model is 2D and covers one ~780 nm field.
- The microstructure run applies the full 1.21 $MW/m^2$, but only ~0.56 $MW/m^2$ crosses the disc center. Absolute fluxes should be scaled by ~0.46.
- The disc uses $k$ = 11.65 W/mK, 0.7 % above the 1500 °C value, which is negligible. Only the product $h(T_\infty - T)$ is constrained at the rim, so σ carries most of the calibration uncertainty.

## Acknowledgements

The ReaxFF MD simulations were performed on the Roar supercomputer managed by the Penn State Institute for Computational and Data Sciences. We thank Texas A&M user facility and A. Bandyopadhyay for the SPS, T. S. Pieshkov, Kevin Beagen, Aidan Donelly, Sean Mcqueen, Melissa Gooch, Attila Thomazy and Bo Chen for their technical helps. We thank Dr. Anahita Emami and Dr. Meysam Khaleghian of Texas State University for providing computational resources including license to ABAQUS/CAE 2024 for finite element analysis. We also thank Robert Purvis, Workshop and Makerspace Director at Rice University for laser engraving. This work is conducted in part using resources of the Shared Equipment Authority at Rice University. X-ray facilities managed by the Shared Equipment Authority of Rice University, USA.

## Funding

P. M. A. acknowledges financial support from the Air Force Office of Scientific Research (AFOSR) under award number FA9550-24-1-0301. P. E. H. acknowledges support from the Office of Naval Research, Grant Number N00014-26-1-2076. A. C. T. van Duin acknowledged funding from ONR/MURI grant number N000142412313. J. H. P. and G. -D. S. acknowledge National Research Foundation of Korea (NRF) Project no: RS-2025-00520263. L. Z. D. acknowledges support from the U.S. Air Force Office of Scientific Research Grants FA9550-15-1-0236 and FA9550-20-1-0068; the T. L. L. Temple Foundation; the John J and Rebecca Moores Endowment; and the State of Texas through the Texas Center for Superconductivity at the University of Houston. L. Z. D. also acknowledges the funding support provided by the Robert A. Welch Foundation #L-E-0001-19921203. TOF-SIMS analysis was carried out with support provided by the National Science Foundation CBET-1626418.

## AUTHOR DECLARATIONS

### Competing interests

P. M. A., A. B., S. S., and E. G. have filed an Intellectual property (IP) disclosure though Rice University Office of Technology Transfer (OTT) with Tech ID No. 2027-031.

### Author contributions

A. B., S. S., and P. M. A. conceptualized the work. A. B., S. C., M. X., and J. L. performed the materials synthesis and characterizations. A. B. and E. G. performed the HVOF tests. M. A. performed the electron microscopy. G. C. performed the EPMA. J. H. P., and G. - D. S. performed the mechanical characterizations. S. D. performed the finite element analysis. M. M., and A. V. D. performed the ReaxFF simulation. T. T. performed the TOF-SIMS. S. Z., J. R., H. B. S., N. B., W. R., M. S. L., and P. E. H., performed the thermal conductivity. M. T., B. K. J., and V. G. performed the tribology. S. K., C. W. C., and L. D. performed the electrical transport. A. B., E. G., K. L., A. V. D., P. E. H., S. S., and P. M. A. analyzed the data. A. B. and P. M. A. wrote the paper with the help of all the authors. All the authors discussed the results and contributed to the manuscript preparation.

### Data availability

All the data are available in the Main text, Methods or in the Supplementary Information. Additional data related to this paper may be requested from the corresponding authors upon reasonable request.

### Additional information

**Supplementary information**
The online version contains supplementary material available at https://doi.org/XXX.

**Correspondence and requests for materials** should be addressed to Abhijit Biswas, Patrick E. Hopkins, Sanjay Sampath, and Pulickel M. Ajayan.

## Supplementary information for

# SiC-BN composites with extreme thermal and environmental stability

## This SI file includes:

Supplementary Figures **1** to **30**

Supplementary Tables **1** to **8**

Supplementary Note

Supplementary Videos **1** to **6**

We provide Supplementary Videos in a separate file.

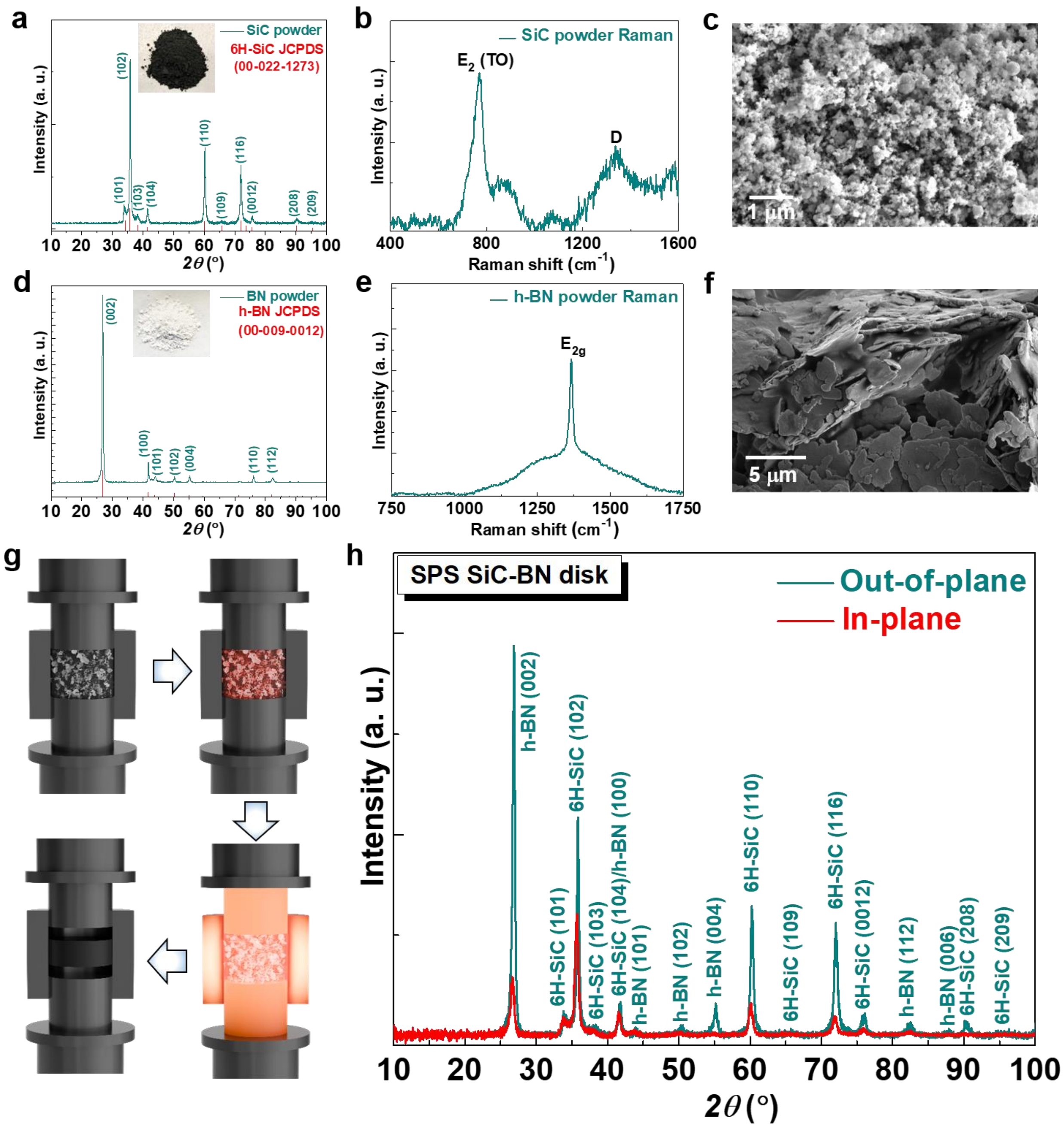

**Fig. S1: Characterizations of pristine 6H-SiC and h-BN powders and the sintered SiC-BN composite. a-f,** X-ray diffraction (XRD), Raman spectroscopy, field-emission scanning electron microscopy (FESEM) characterizations of the pristine 6H-SiC and h-BN powders showing the 3D particle and 2D sheet-like morphology. **g,** Illustration of the spark plasma sintering (SPS) process used to fabricate the SiC-BN composite disk. **h,** Out-of-plane and in-plane XRD patterns of the sintered SiC-BN disk, confirming the crystallographic orientation of the constituent phases.

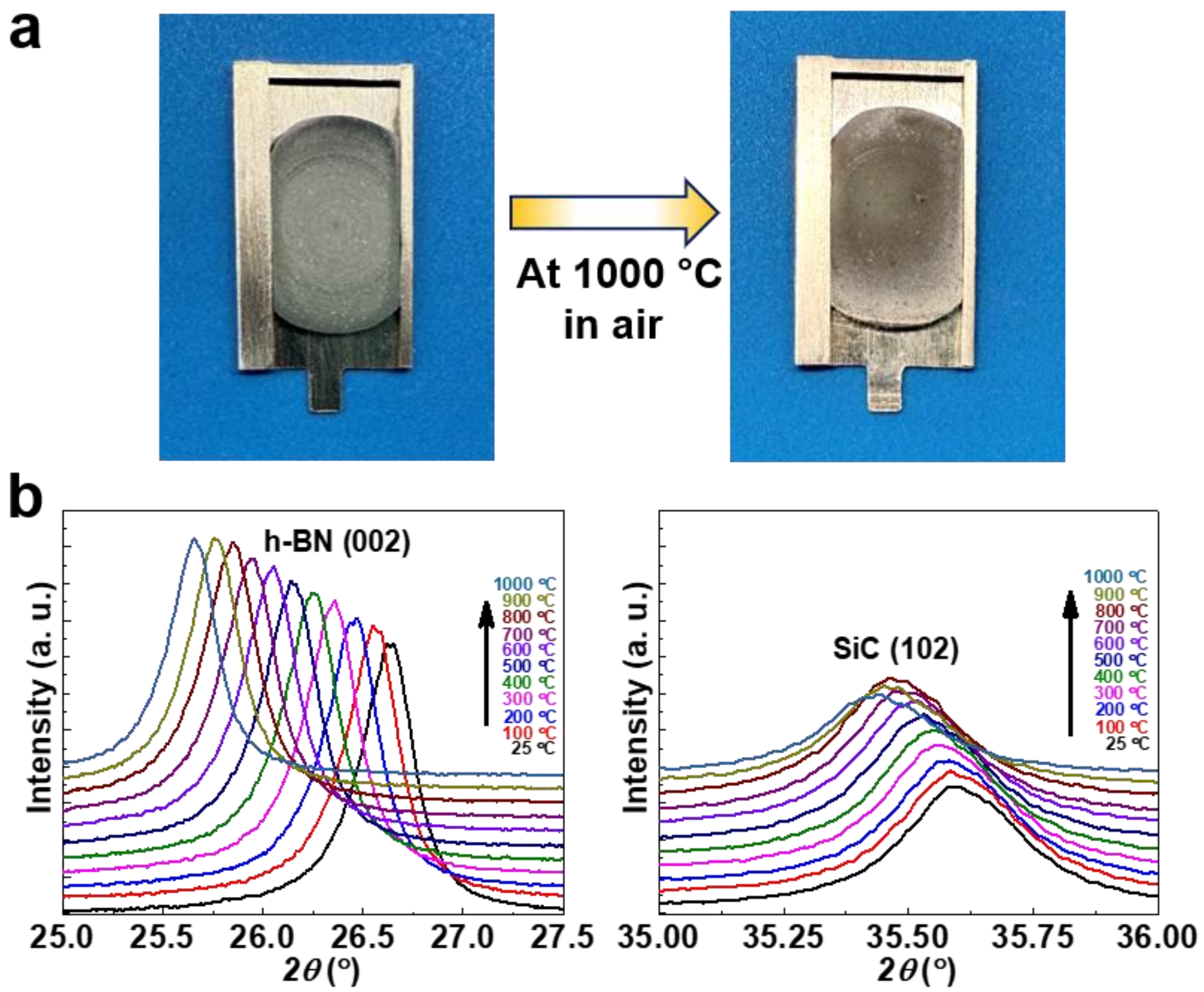

**Fig. S2: Temperature-dependent X-ray diffraction of the composite. a,** Image of the composite after the XRD in air up to 1000 °C (the instrumental limit). **b,** Temperature-dependent XRD patterns revealing the lattice thermal expansion of the respective phases.

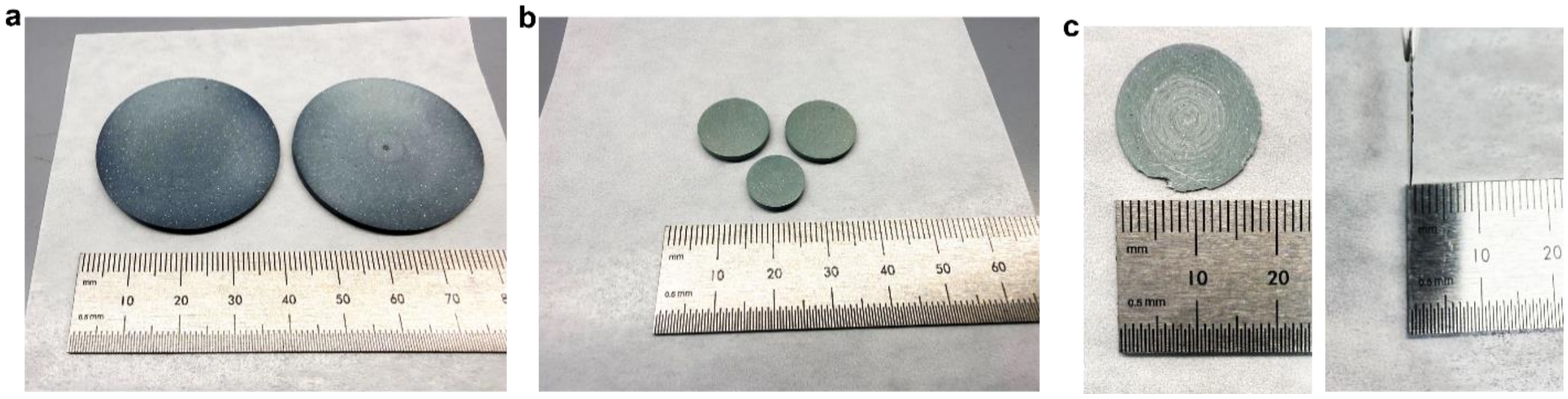

**Fig. S3: Machinability of the composite. a, b,** Composites are easily machinable (by using a diamond blade), enabling a thick disk sliced into thinner or smaller-diameter disks. **c,** It can be sliced down to ~200 μm thin disk without breaking.

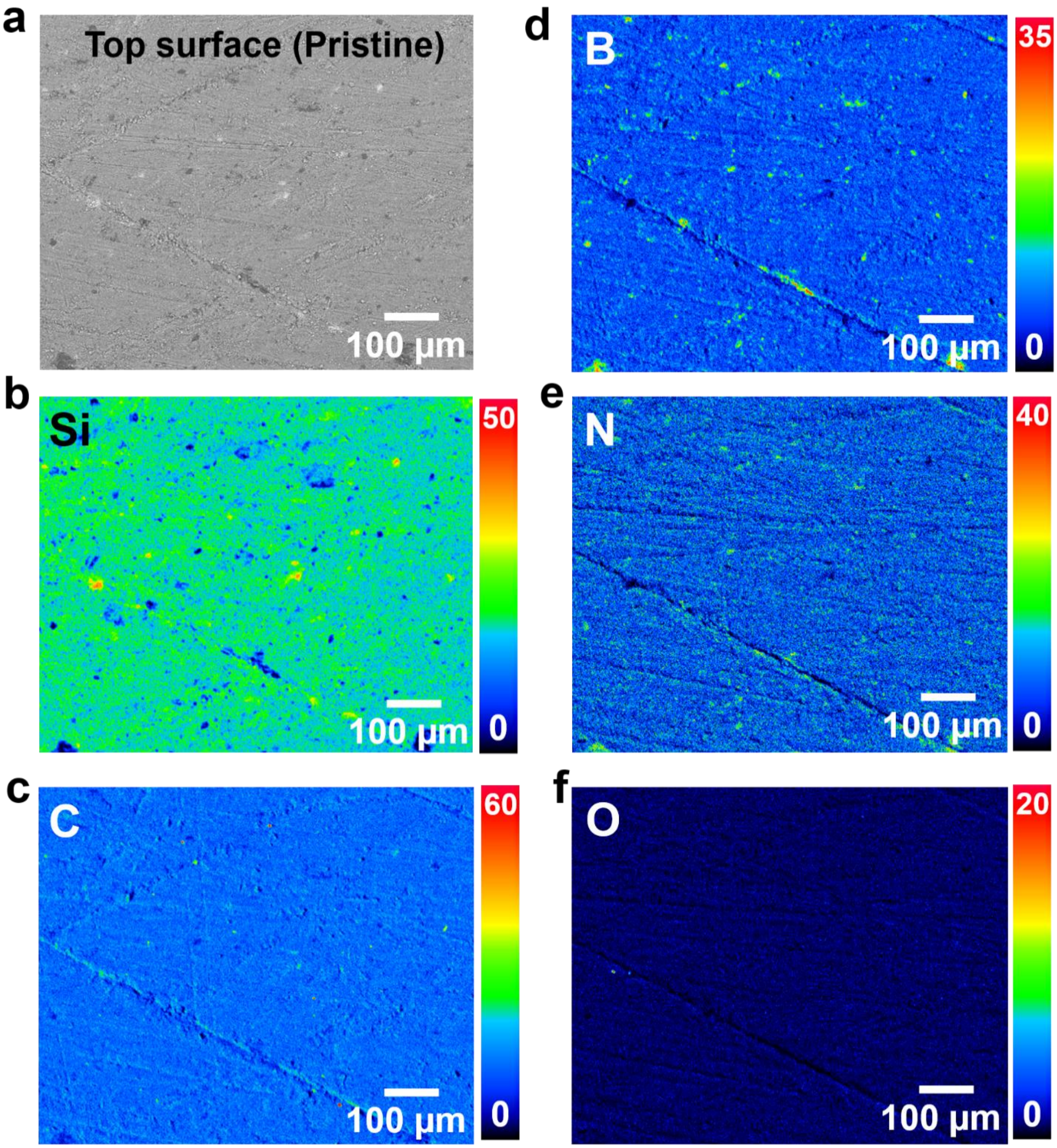

**Fig. S4: Electron probe microanalysis (EPMA) compositional map of pristine composite. a-f**, EPMA backscattered electron image and corresponding wavelength-dispersive spectrometry (WDS) compositional maps showing the presence of elements with an wt% of 32.93% (Si), 24.38% (C), 12.53% (B), 27.89 %(N), and 2.34 %(O).

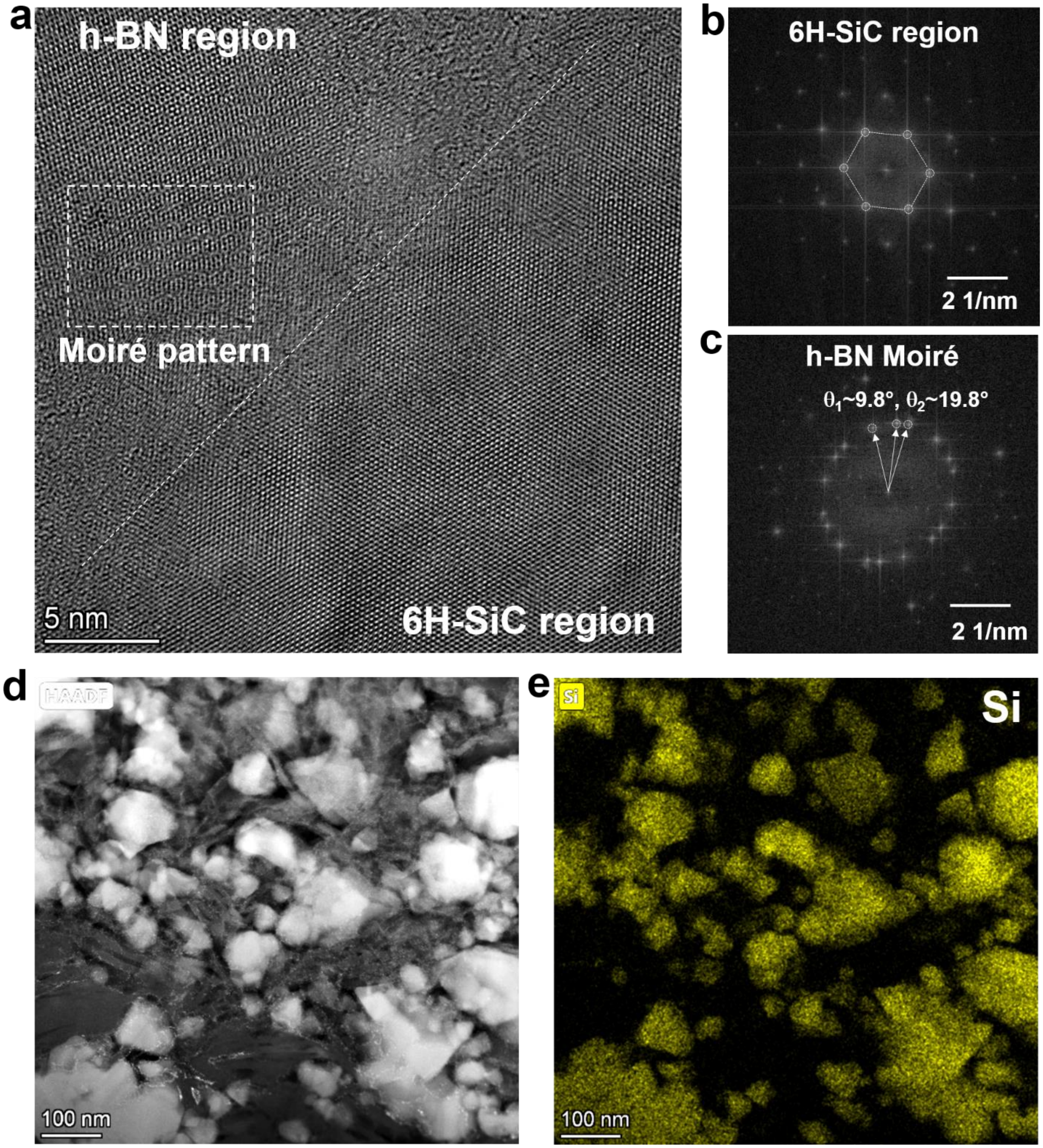

**Fig. S5: Atomic-scale characterization of the composite. a, b, c,** HRTEM image from a different region of the composite showing hexagonal diffraction pattern from the 6H-SiC region and Moiré patterns from the h-BN region (similar like **Fig. 1f-h**). **d, e,** The HAADF-STEM microstructural image and corresponding energy-dispersive X-ray spectroscopy (EDS) elemental map showing the distribution of Si (due to low *Z* number, B and N were difficult to map).

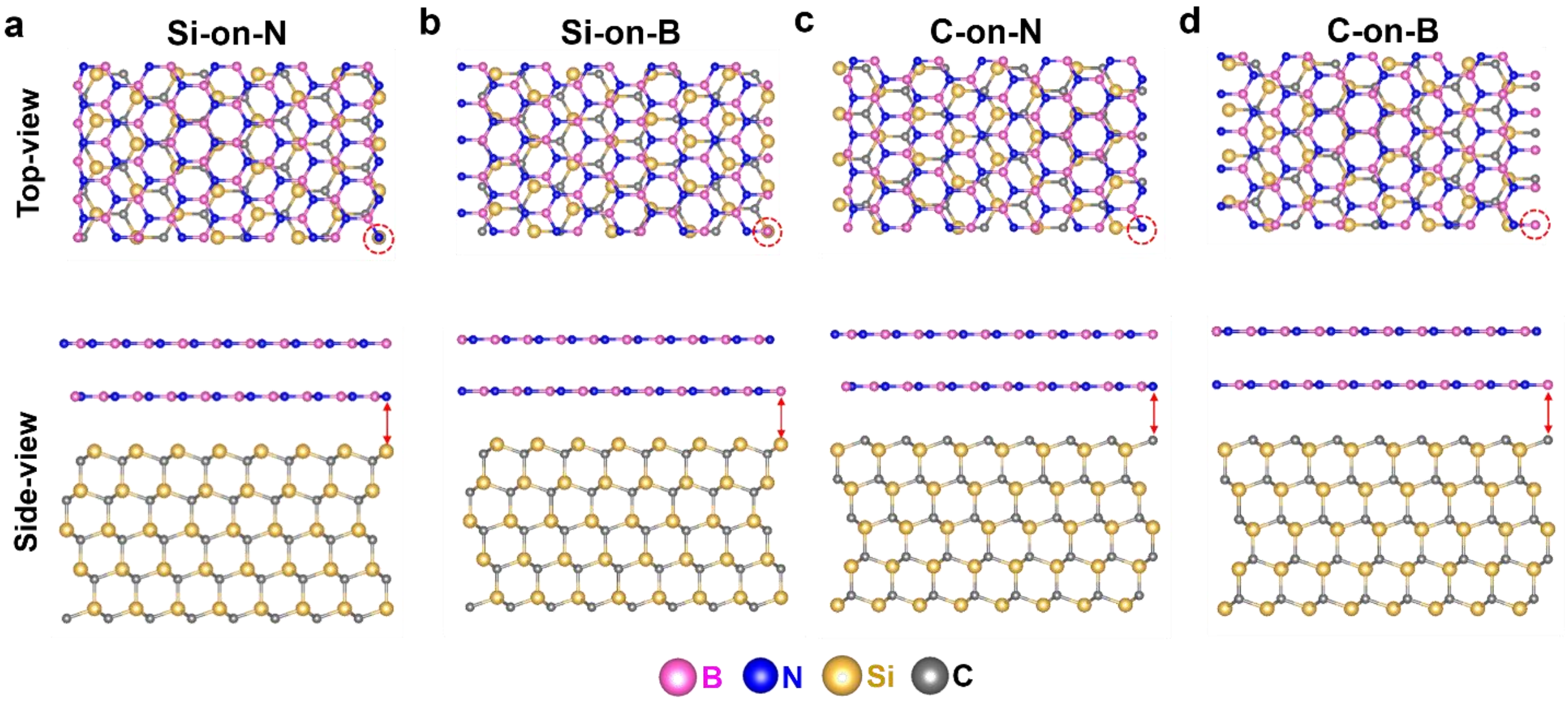

**Fig. S6: Top and side-views of the 6H-SiC/h-BN interfaces.** The columns correspond to the on-top-of-N and on-top-of-B interface configurations, respectively. **a, b,** show the Si-terminated interfaces, while **c, d,** show the C-terminated interfaces. For the top views, only the topmost SiC bilayer and bottommost h-BN layer are shown for clarity. The red arrows and dotted circles mark the reference (0,0) position, indicating whether the surface Si (or C) atom is aligned beneath an N or B atom for the on-top-of-N and on-top-of-B configurations, respectively.

**Table S1: Calculated interface formation energies of the SiC-BN heterostructures (unit: Kcal/mol·Å$^2$).**

| Interface configuration | Si-terminated | C-terminated |
|---|---|---|
| On top of N | -1.051549 | -0.383307 |
| On top of B | -1.046537 | -0.383351 |

The relative stability of the different SiC/BN interface configurations was evaluated using the interface formation energy:

$$\gamma_{int} = \frac{E_{SiC/BN} - E_{SiC} - E_{BN}}{A}$$

Where $E_{SiC/BN}$ is the energy of the relaxed SiC/h-BN heterostructure, $E_{SiC}$ and $E_{BN}$ are the energies of the corresponding relaxed isolated SiC and h-BN slabs with the same surface area and computational conditions, and A is the interfacial area. More negative values of $\gamma_{int}$ indicate a more energetically favorable interface according to this definition. For both Si- and C-terminated surfaces, the on-top-of-N configuration exhibits a lower interface energy than the corresponding on-top-of-B configuration.

The preference for N-site alignment is consistent with the greater electrostatic attraction expected between the relatively electropositive surface sites of SiC and the electronegative N atoms in h-BN. In contrast, the on-top-of-B configurations are less favorable because of less favorable local chemical and electrostatic interactions at the interface. Among the configurations examined, the Si-terminated/on-top-of-N interface exhibits the lowest calculated interface energy and was therefore identified as the most stable configuration within the present ReaxFF model.

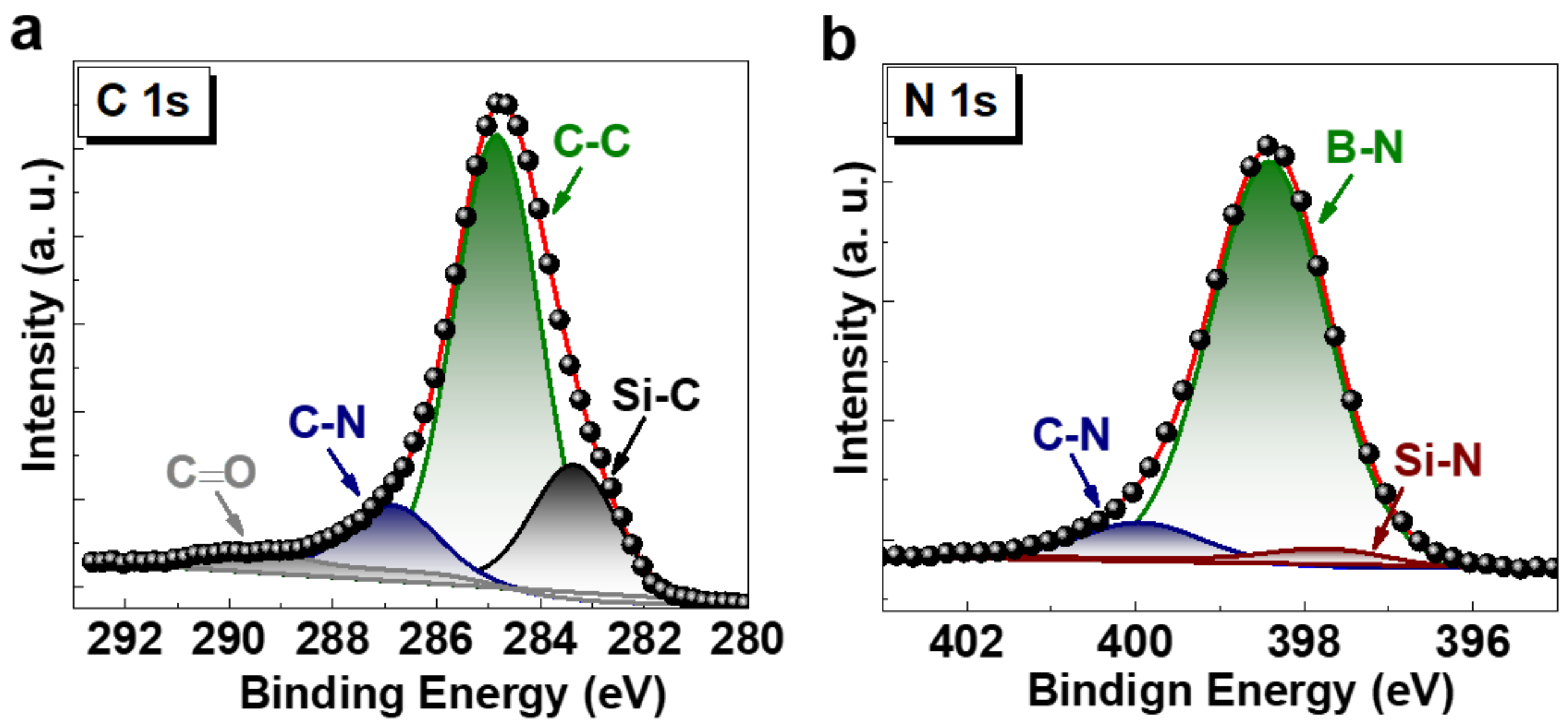

**Fig. S7: Elemental XPS analysis of the composite.** XPS spectra of **a,** C 1s, and **b,** N 1s core level also reveals the characteristic Si-C and B-N bonding environments, together with additional spectral features associated with interfacial bonding between the SiC and BN phases.

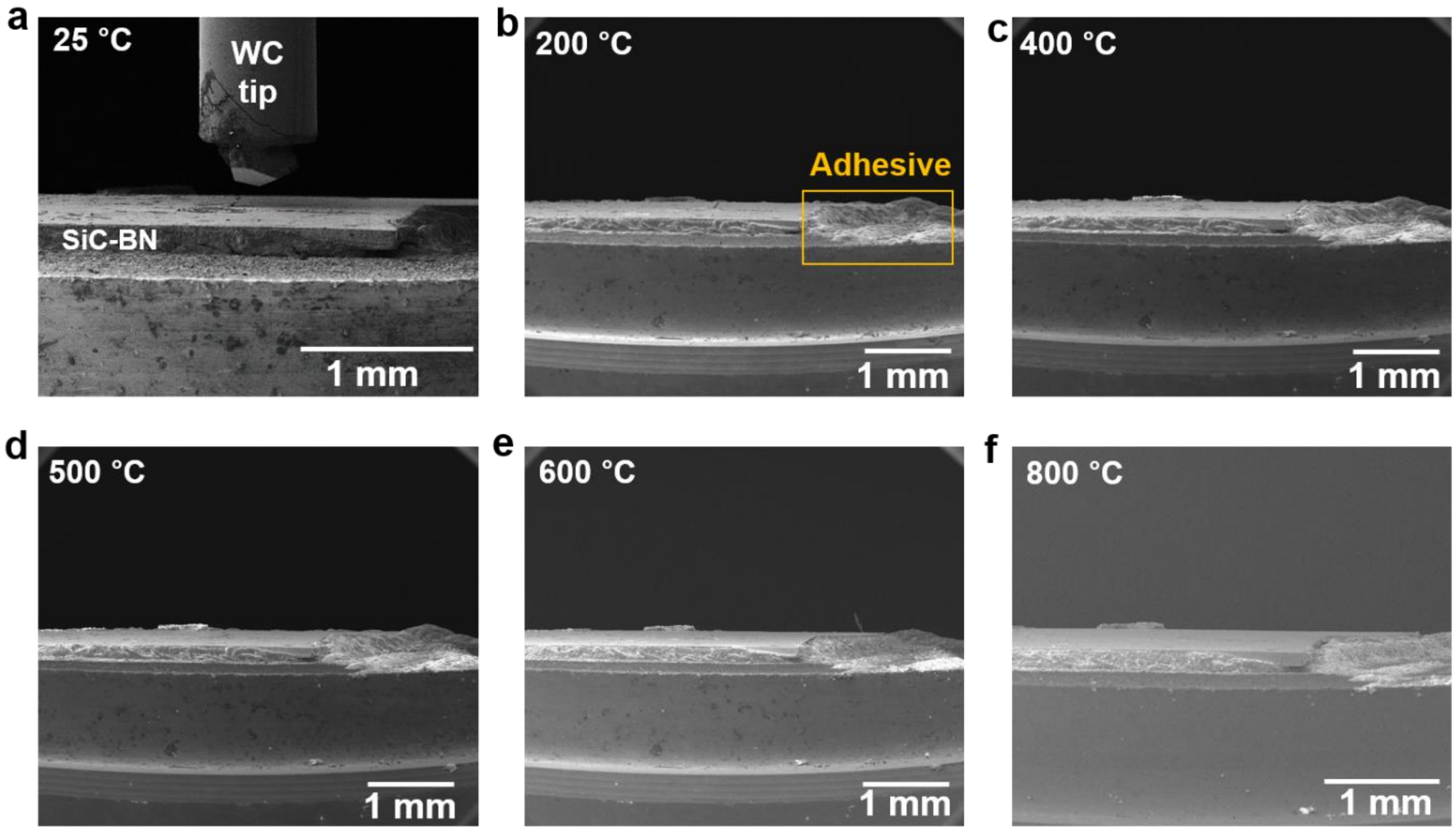

**Fig. S8: Temperature-dependent nanoindentation of the composite. a-f,** SEM images of the SiC-BN slab glued to the indentation stub with load applied through tungsten carbide (WC) Berkovich tip. Images were taken while the sample stage temperature was raised from 25 to 800 °C.

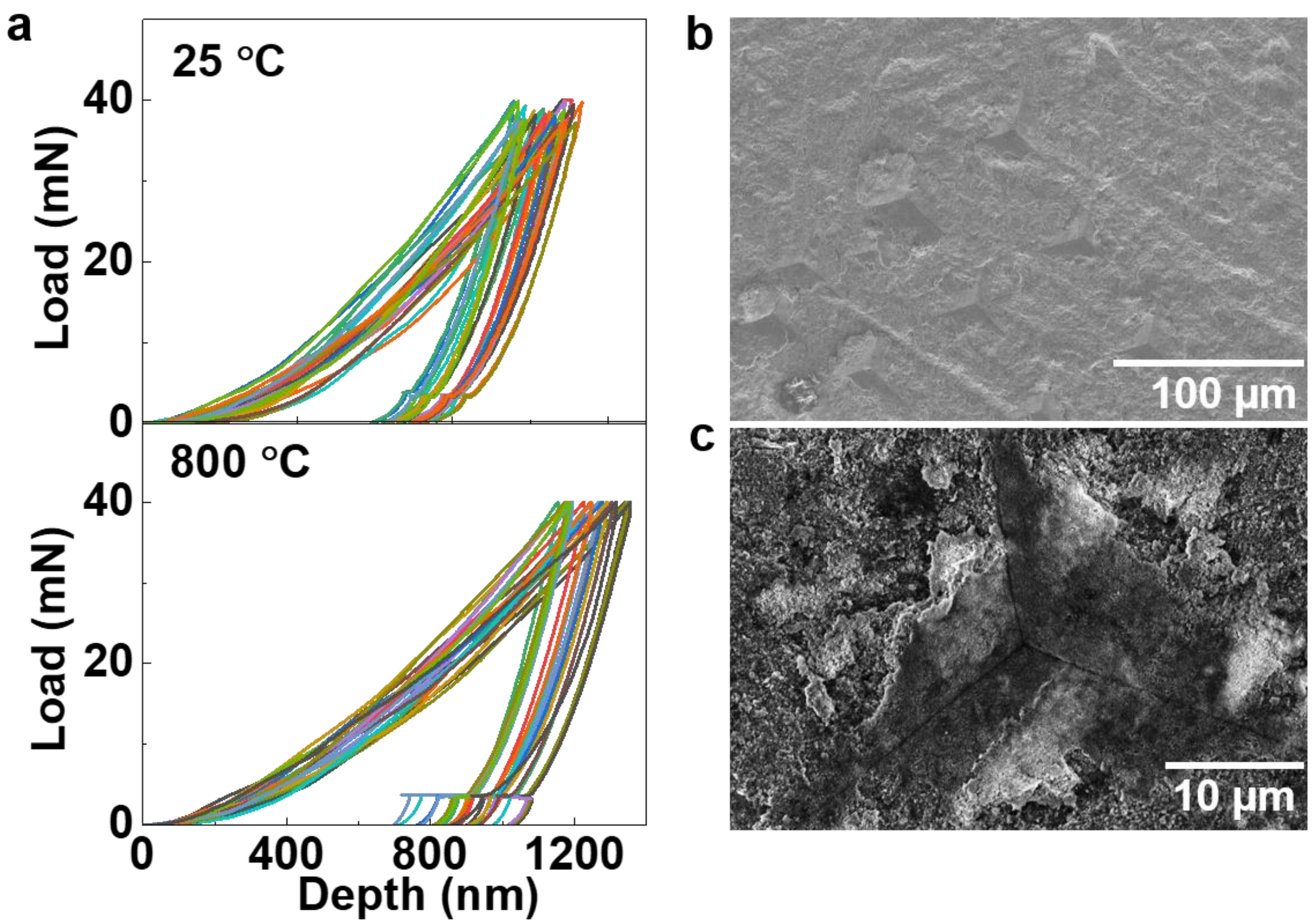

**Fig. S9: Nanoindentation test of the composite. a,** Load vs. displacement curves at 25 and 800 °C. **b,** Large-scale low-magnification SEM image of the indented surface with a Berkovich tip at a peak load of *P* = 1000 mN, showing an array of nanoindentation spots. **c,** Higher-magnification image revealing the features of the individual indentation region.

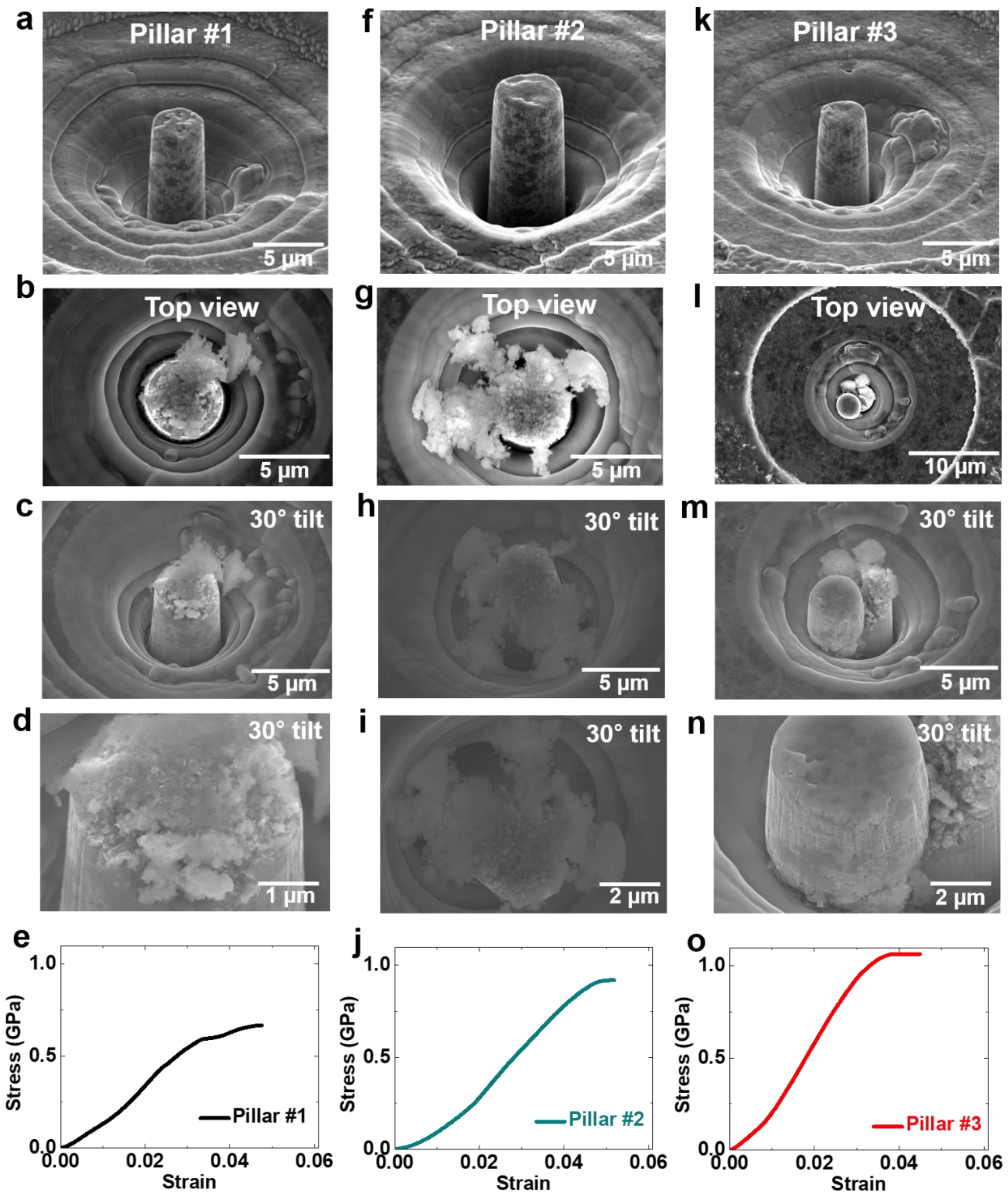

**Fig. S10: Micropillar compression test of the composite.** Fabricated three micropillars using focused ion beam (FIB) milling inside the SEM and post-mortem SEM images of the pillars showing the fracture. All pillars show features of brittle fracture. **a-d,** Pillars #1, and **e-f**, #2 completely crumbled after compression, while the part that was in contact with the flat punch tip is still visible in **i-l,** pillar #3. **m,** Stress-strain curve obtained at 25 °C, showing the yield strength and compressive strength of the composite.

**Table S2: Mechanical properties of the SiC-BN micropillars.**

| Pillar No | Top pillar diameters (μm) | Bottom pillar diameters (μm) | Pillar height (μm) | Taper angle (°) | Elastic modulus (GPa) | Yield strength (MPa) | Failure strength (MPa) |
|---|---|---|---|---|---|---|---|
| 1 | 3.66 | 4.33 | 10.51 | 2.20 | 25.0 | 612 | 667 |
| 2 | 3.58 | 4.19 | 8.91 | 1.93 | 27.2 | 908 | 919 |
| 3 | 3.19 | 3.81 | 8.00 | 2.22 | 37.5 | 1064 | 1066 |

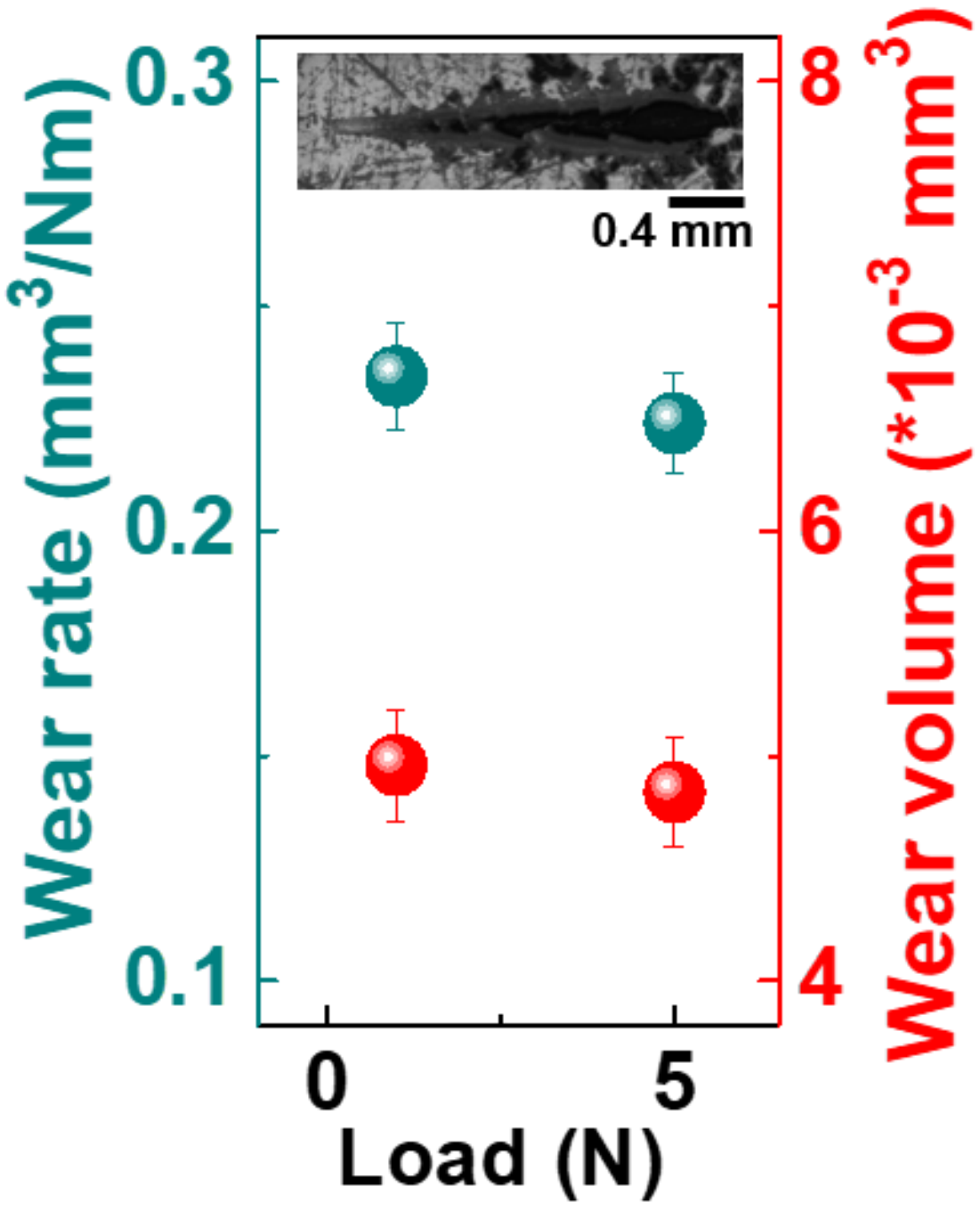

**Fig. S11: Reciprocal wear-test of the composite.** Room-temperature wear map (inset) and corresponding wear rate and volume obtained from a linear reciprocating ball-on-flat-disk wear test using a zirconia ball.

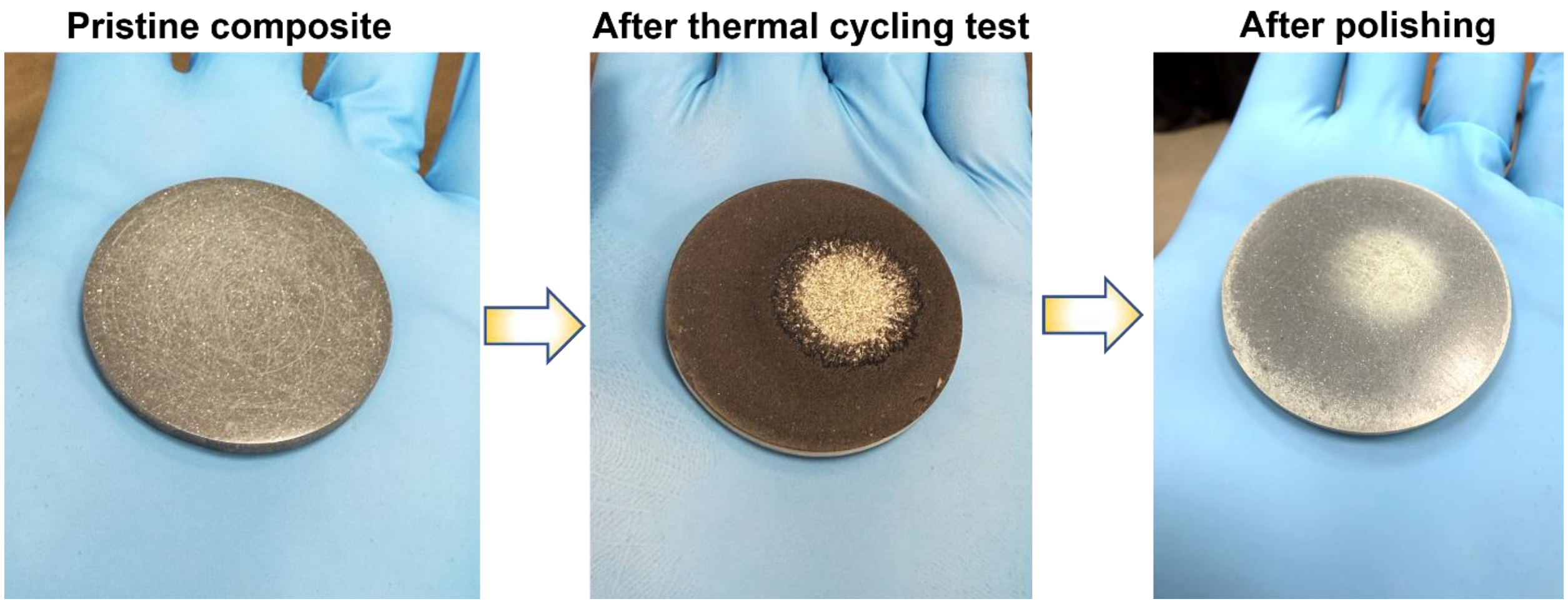

**Fig. S12: Reusability of the composite disk after the HVOF test.** After thermal-cycling test the disk surface was mechanically polished with sand paper and ready for reused for the future experiments.

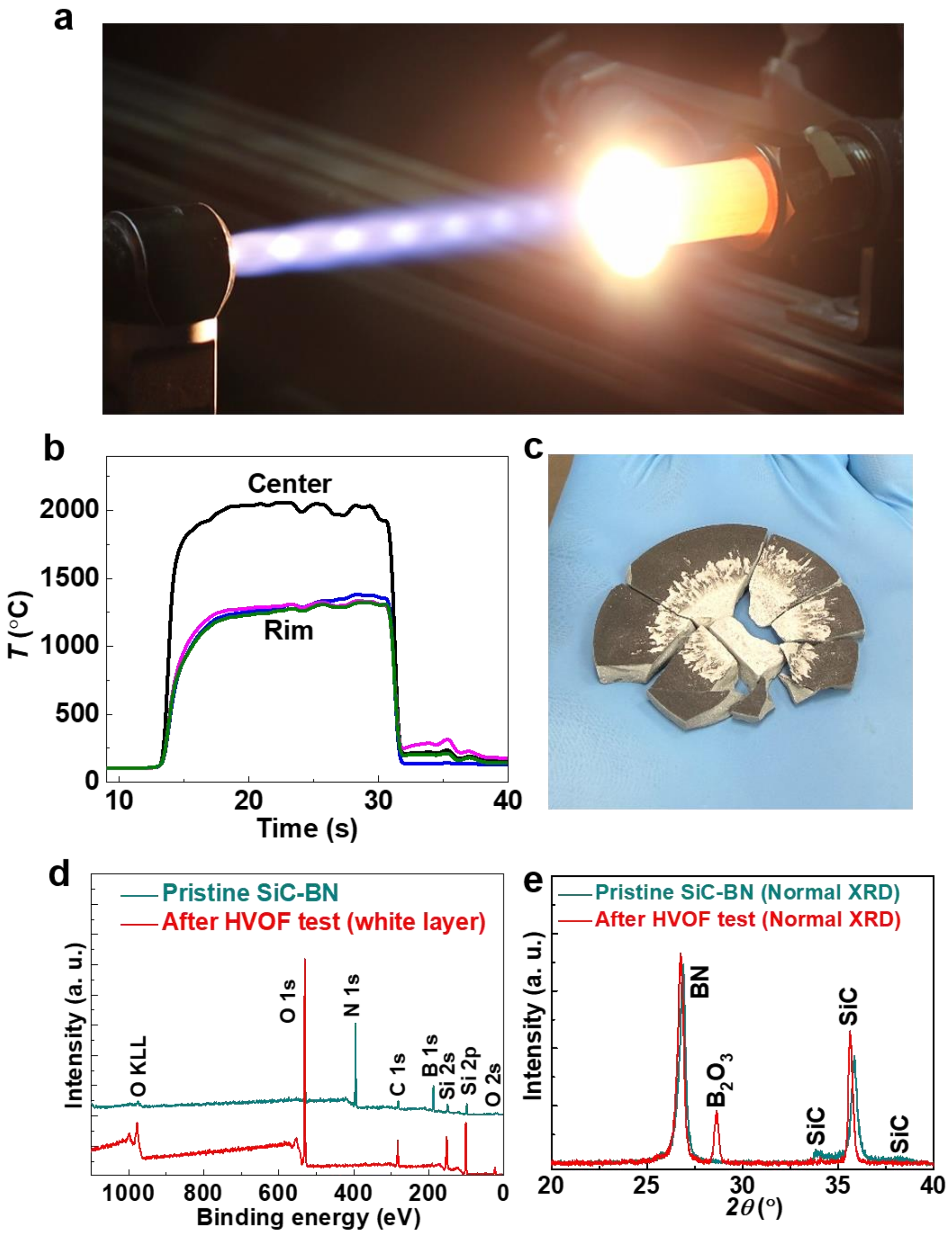

**Fig. S13: HVOF torch impact test at 2000 °C of the composite. a-c,** HVOF torch impact test on the SiC-BN disk at a front-face surface temperature of ~2000 °C showing that the disk maintains structure only for ~20 s before fracture. The front-face temperature was increased to ~2000 °C by reducing the torch stand-off distance to 15 cm, compared to the 20 cm distance used to achieve ~1500 °C, **Fig. 3a**). **d, e,** XPS analysis of the white surface layer showing the presence of Si and O, and bulk XRD revealing the presence of $B_2O_3$ at the white spot.

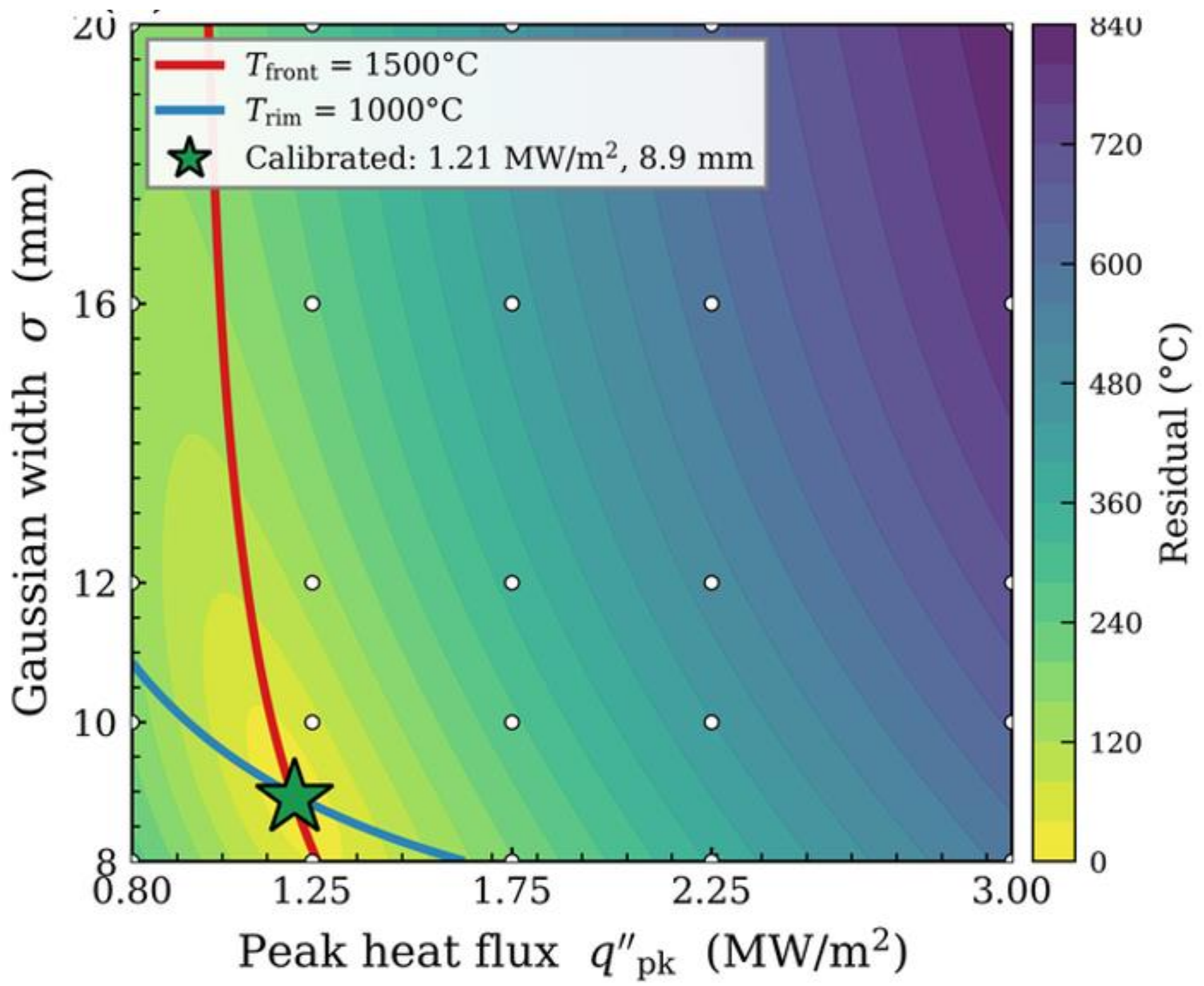

**Fig. S14: Inverse Calibration.** Residual between simulated and measured temperatures across the 25-run sweep. The red and blue curves are the $T_1$ =1500 °C and $T_3$ =1000 °C iso-contours, and the star marks the calibrated point.

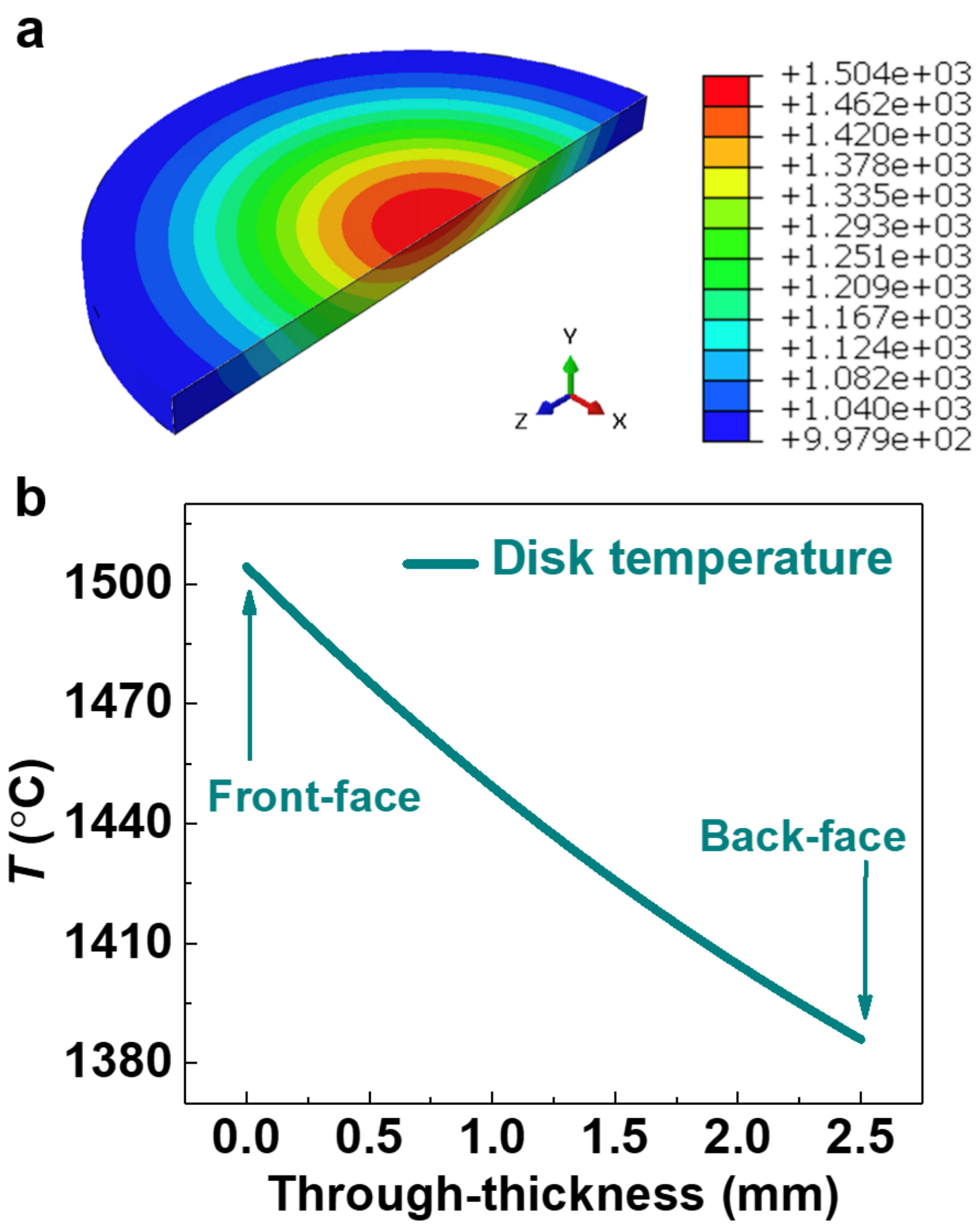

**Fig. S15: Temperature distribution within the composite. a, b,** Calculated Finite element analysis calculated temperature distribution though the 2.5 mm thick disk during the HVOF test.

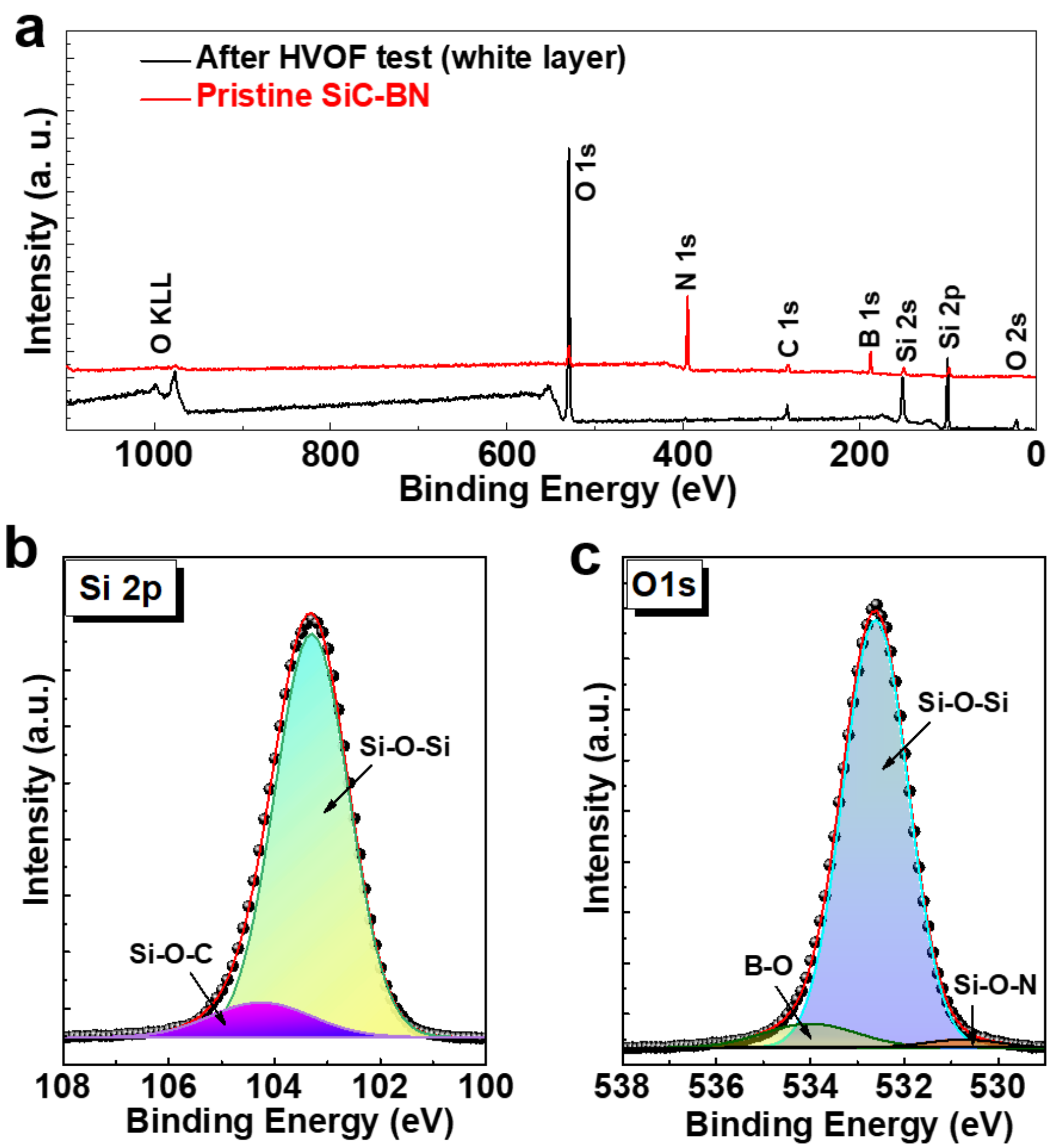

**Fig. S16: Elemental XPS analysis of the post-impacted surface (top white layer). a,** XPS survey spectrum of the white surface layer reveals Si and O as the predominant elements. **b, c,** XPS spectra on the white layer shows the formation of predominant Si-O-Si, together with a smaller fraction of other bonding's.

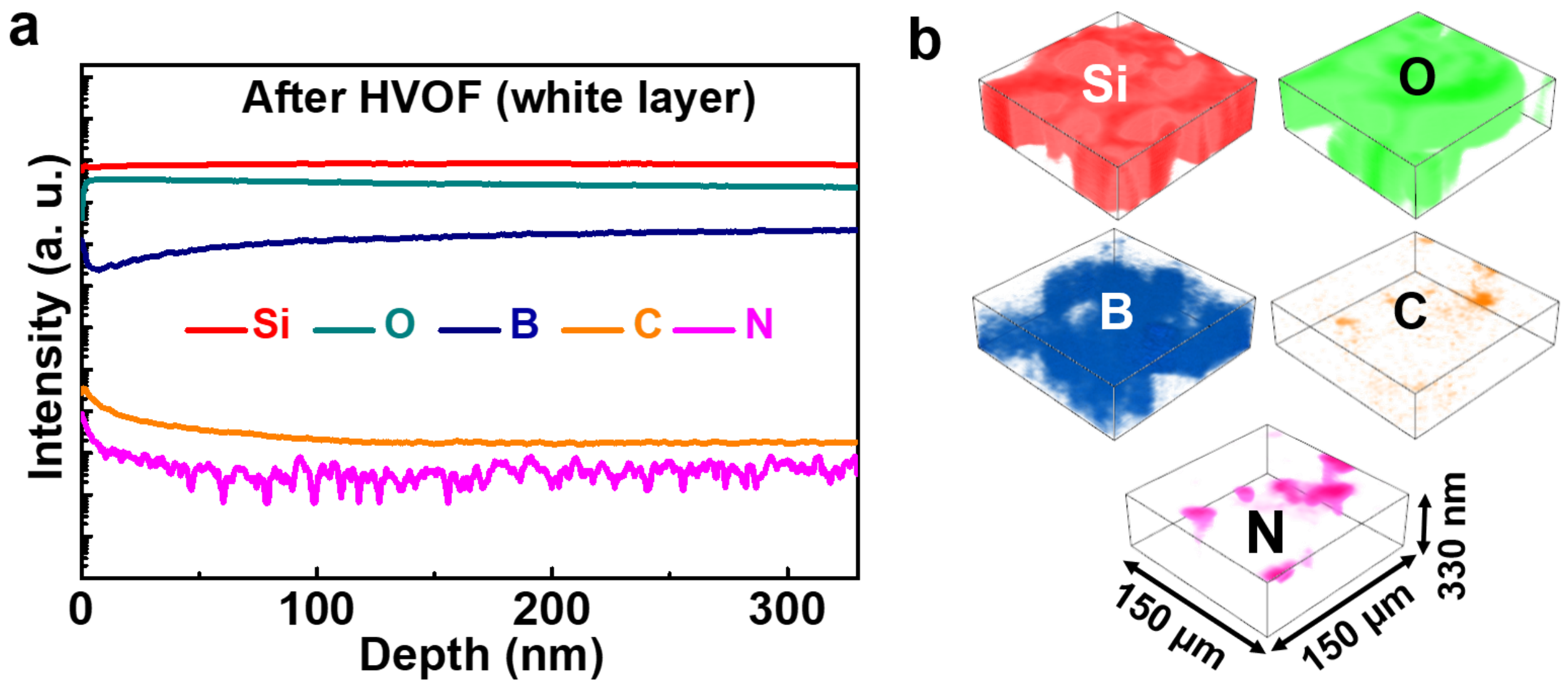

**Fig. S17: TOF-SIMS of the white surface layer after the HVOF test. a, b,** TOF-SIMS depth profile and corresponding elemental maps show the high presence of Si, and O within the white surface layer, with negligible signal of B, C, and N at the surface. Boron becomes detectable ~10-20 nm beneath the surface, indicating a near-surface depletion of B following HVOF exposure. This is consistent with the high-temperature volatilization of boron oxide species, including $B_2O_3$, during oxidation at ≥800 °C.

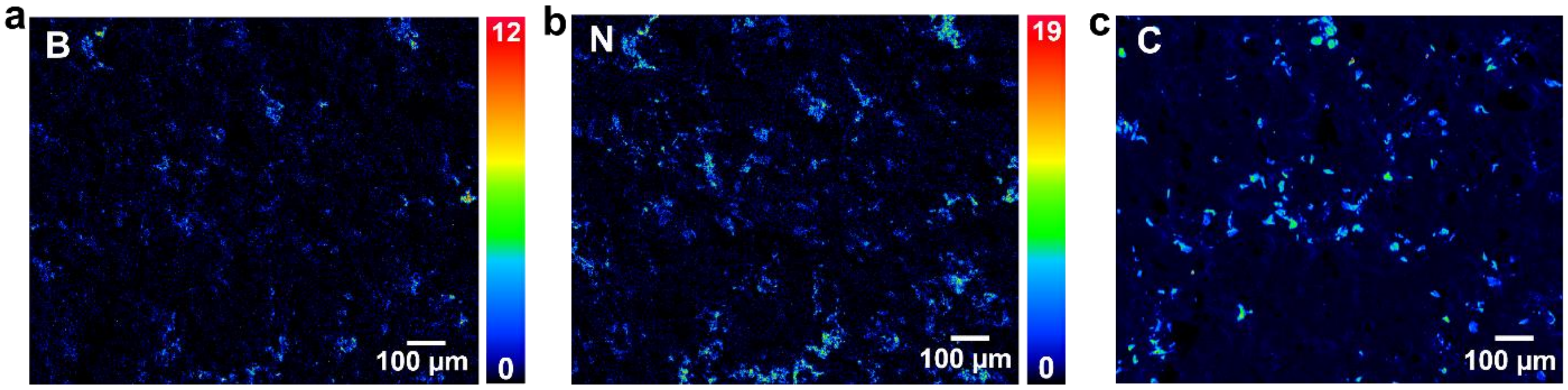

**Fig. S18: EPMA elemental compositional analysis of the top surface white layer after the HVOF impact. a-c,** EMPA WDS elemental maps of B, N, and C on the top surface (white layer) with an obtained wt% of 0.65% (B), 15.18% (N), and 6.81% (C).

**Table S3: Post-impact compositional analysis (wt%) of the SiC-BN composite.**

| **Analyzed area** | **No. of points** | **Si (wt.%)** | **O (wt.%)** | **C* (wt.%)** | **B (wt.%)** | **N (wt.%)** |
|---|---|---|---|---|---|---|
| White layer **(top surface)** | 52 | 28.03 | 48.38 | 6.81 | 0.65 | 15.18 |
| Black layer **(top surface)** | 110 | 25.92 | 48.66 | 9.10 | 0.47 | 15.25 |
| White layer **(cross-section)** | 18 | 41.80 | 20.21 | 3.50 | 11.24 | 24.20 |
| Below the white layer **(cross-section)** | 25 | 29.77 | 2.45 | 24.37 | 13.83 | 30.87 |
| Pristine sample (**top surface**) | 25 | 32.93 | 2.34 | 24.38 | 12.56 | 27.89 |

*C was corrected for carbon coating and carbon deposition

**Table S4: Post-impact compositional analysis (At%) of the SiC-BN composite.**

| Analyzed area | Chemical formula | Si (At%) | O (At%) | C (At%) | B (At%) | N (At%) |
|---|---|---|---|---|---|---|
| White layer **(top surface)** | $SiO_3C_{0.57}N_{0.57}B_{0.05}$ | 19.32 | 57.68 | 11.09 | 1.00 | 10.92 |
| Black layer **(top surface)** | $SiO_{3.2}C_{0.78}N_{0.57}B_{0.05}$ | 17.88 | 57.14 | 13.98 | 0.82 | 10.18 |
| White layer **(cross-section)** | $SiO_{0.85}C_{0.2}B_{0.67}N_{1.1}$ | 26.08 | 22.15 | 5.16 | 17.40 | 29.20 |
| Below the white layer* **(cross-section)** | $Si_4C_7B_4N_7$ | 15.86 | 2.29 | 30.22 | 18.90 | 32.73 |
| Pristine sample* (**top surface**) | $Si_4C_7B_4N_7$ | 18.09 | 2.28 | 30.93 | 17.70 | 31.01 |

*For last two cases, it falls in the compositional range of SiC-BN ceramic with only minor oxygen presence which shows up during the EPMA sample preparation.

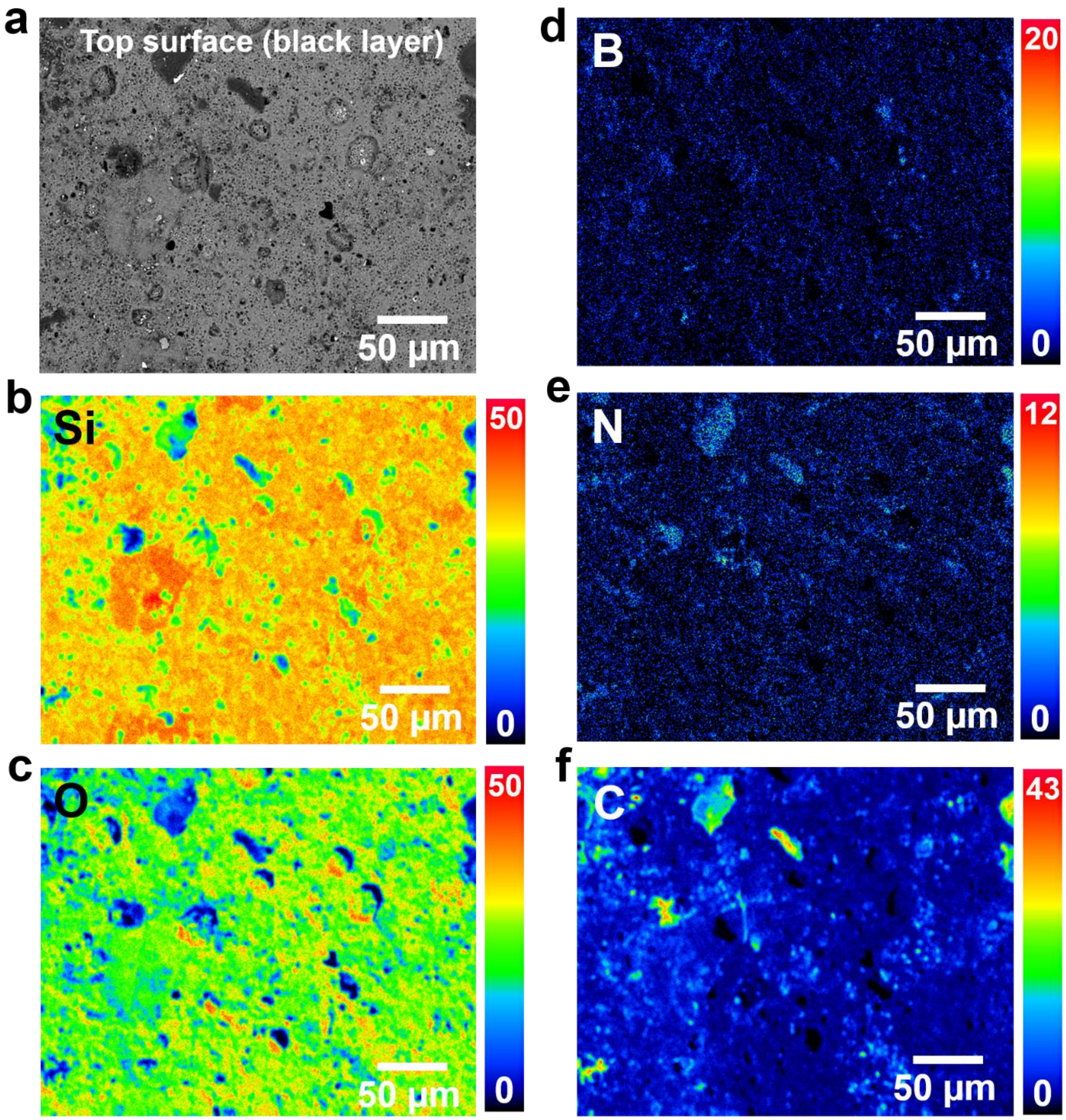

**Fig. S19: EPMA elemental compositional analysis of the top surface black layer after the HVOF impact. a-f,** The EPMA-BSE image and compositional map from the black layer (top surface) shows the presence of elements with an obtained wt% of 25.92% (Si), 48.66 % (O), 15.25% (N), 9.10% (C), and 0.47% (B).

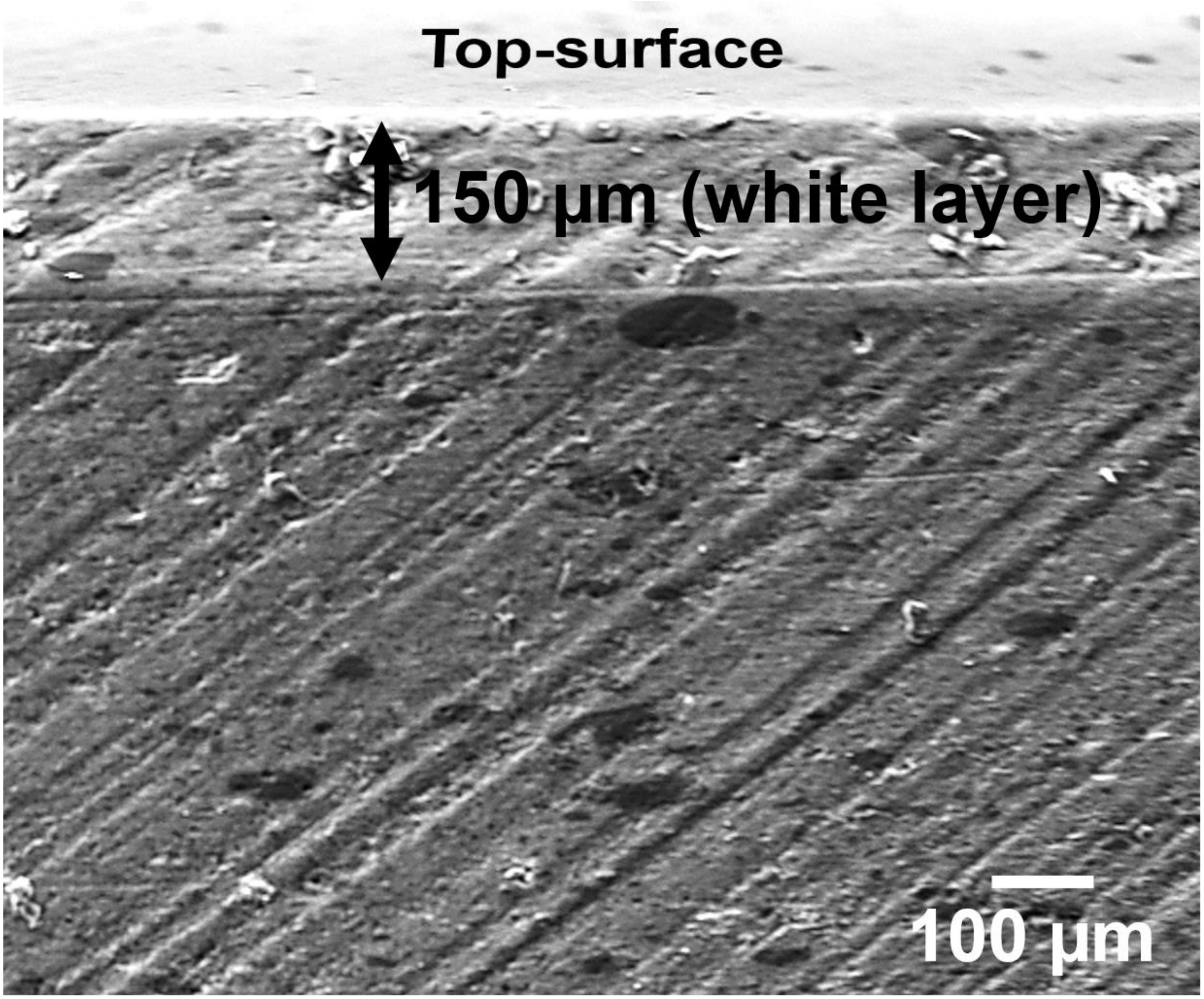

**Fig. S20: Cross-sectional image of the composite after the HVOF impact test.** Cross-sectional EPMA secondary-electron image shows two distinct layers across the disk thickness, with the white layer of ~150 μm into the material. The diagonal features arise from the polishing process.

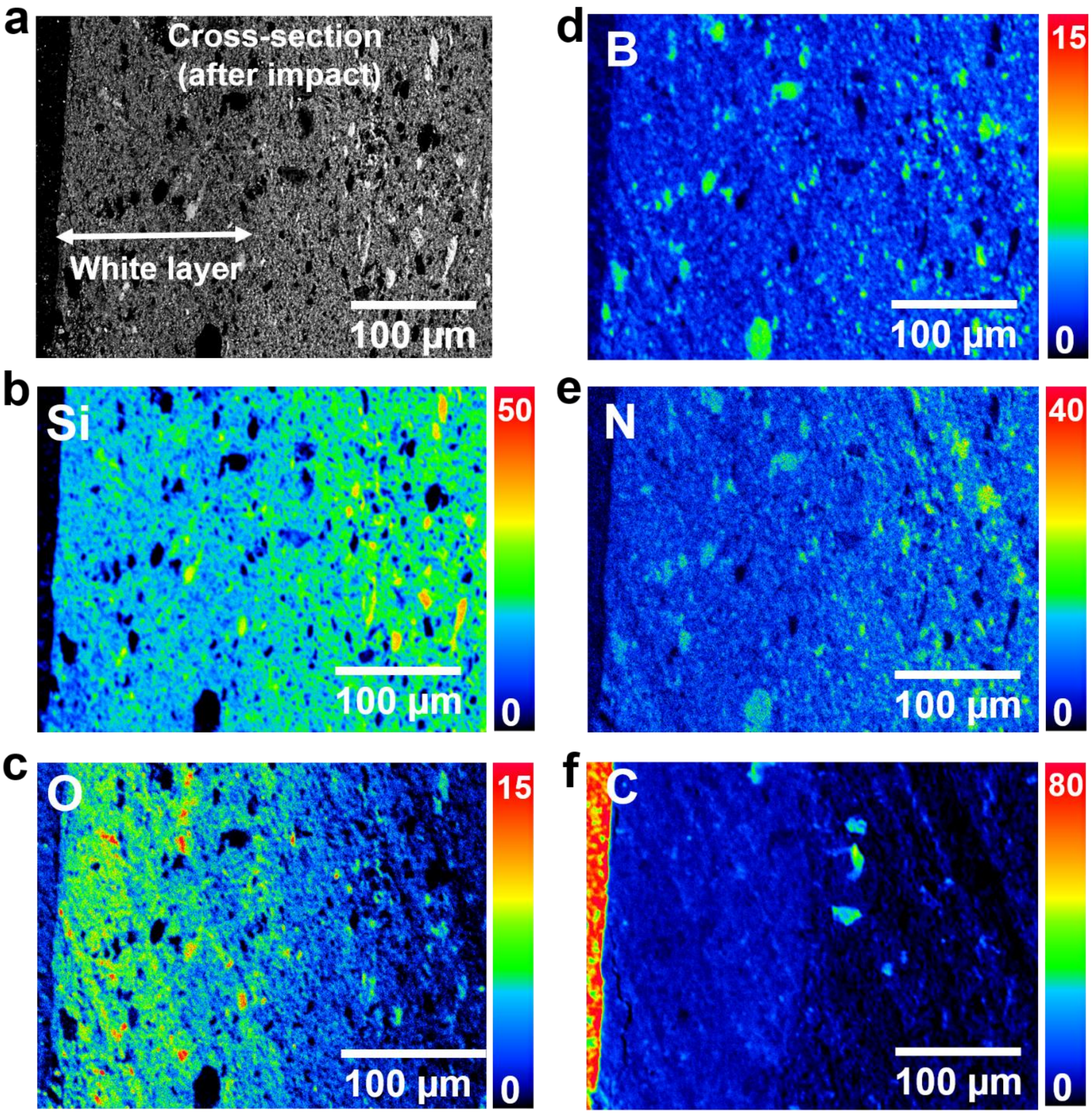

**Fig. S21: Cross-sectional EPMA and elemental compositional analysis of the after-impacted composite. a-f,** The EPMA-BSE image and compositional map of the white (cross-section) layer (~150 µm) shows the presence of elements with an obtained wt% of 41.08% (Si), 20.21% (O), 24.20% (N), 3.50% (C), and 11.24% (B), showing drop in oxygen and increase of Si, B, and N signal. The compositional map below the white 150 µm layer (cross-section) shows the presence of elements with an obtained wt% of 29.77% (Si), 2.45 % (O), 30.87% (N), 24.37% (C), and 13.83% (B), similar to the pristine SiC-BN.

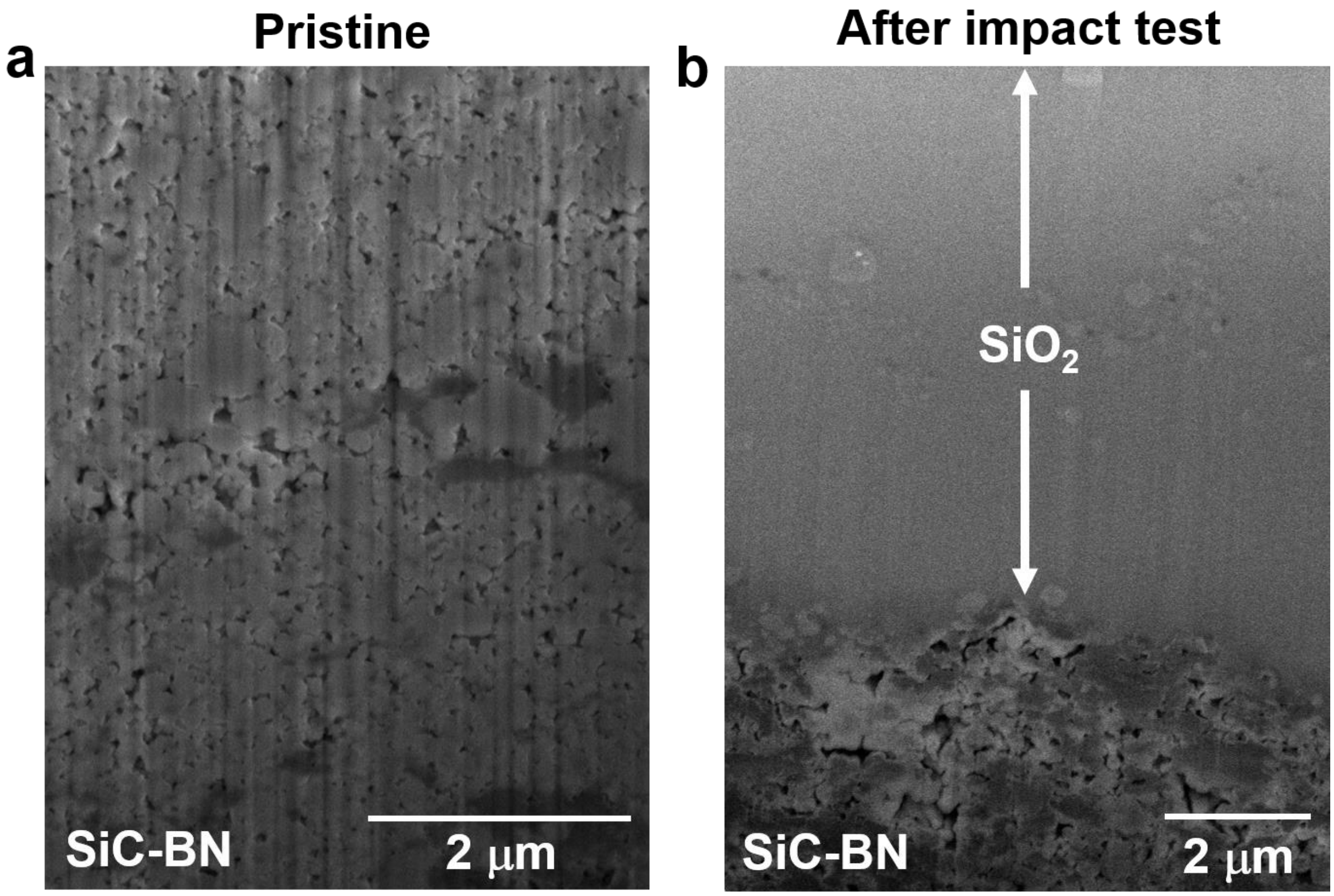

**Fig. S22: Morphological comparison of the pristine and impacted composite. a,** Cross-sectional FESEM image of the pristine composite showing the microstructure. **b,** Cross-sectional FESEM image of the composite after the HVOF impact test. Here we only shown the interfacial region (with an overlying oxide layer). Despite the formation of the oxide layer, the underlying SiC-BN composite retains a microstructure comparable to that of the pristine SiC-BN (also quantitively confirmed by the EPMA analysis, **Table S4**).

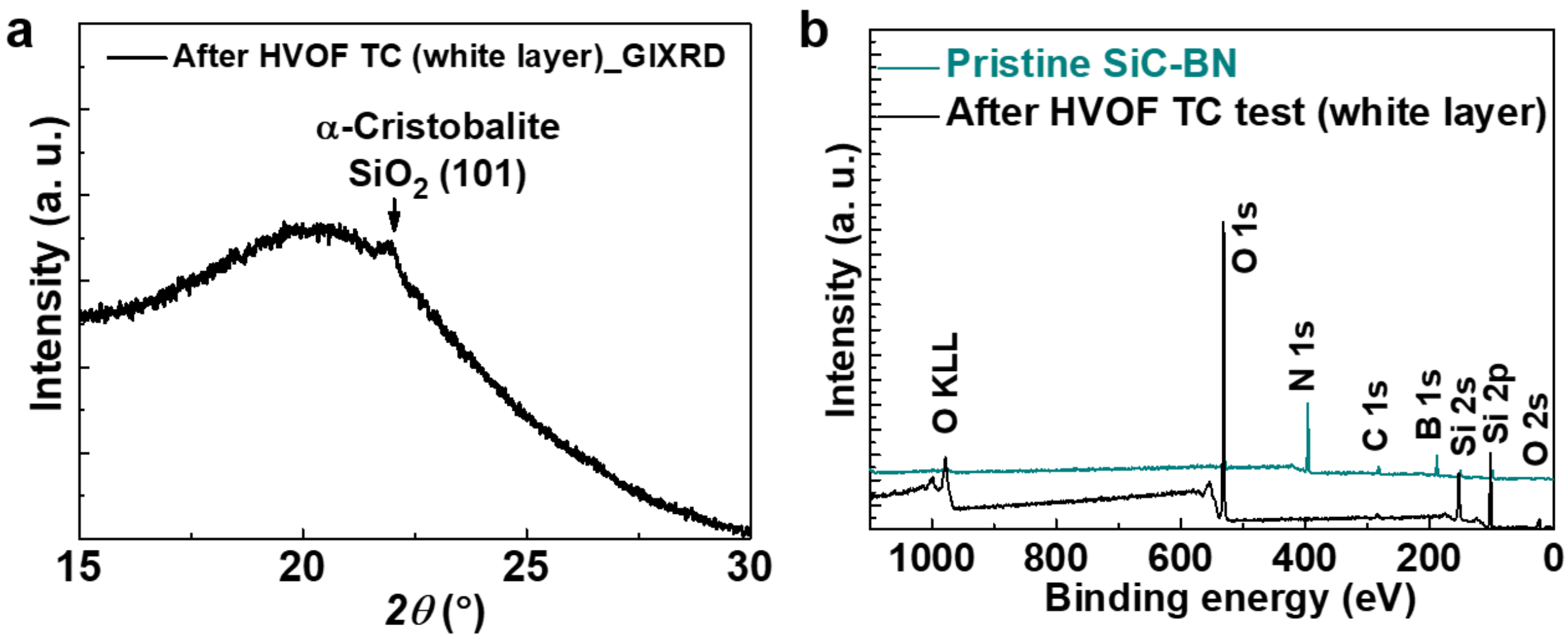

**Fig. S23: Characterizations of the composite after HVOF thermal cycling (TC) test. a,** Grazing-incident angle x-ray diffraction (GIXRD) from the white surface layer after the thermal cycling test (6 min per cycle with 10 s torch-off between each cycle, with three heating-cooling cycles) shows both amorphous and cristobalite $SiO_2$ phases. **b,** XPS survey spectrum of the white layer showing predominantly Si and O, with no detectable B or N. From XPS, the obtained At% of the elements are Si (29.6), O (69.0), and C (1.4), respectively.

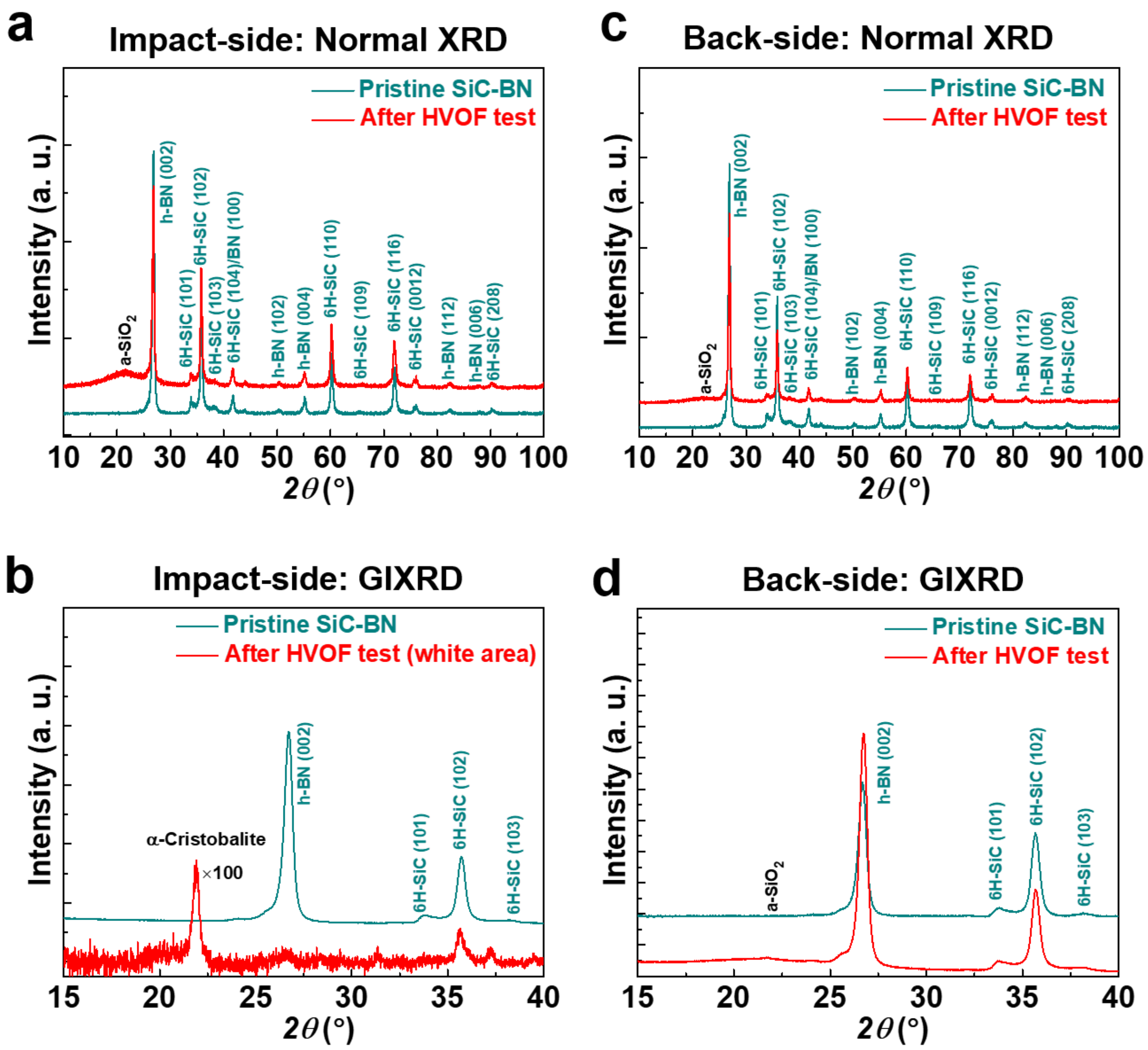

**Fig. S24: XRD comparison of the pristine composite and after the HVOF test. a, b,** Bulk XRD patterns show that the disk retains its 6H-SiC and h-BN phases, even after the 6 min impact test, with a small amorphous silica (a broad hump around 20-25°) peak, and a $\alpha$-cristobalite silica peak on the impact-side top-surface (appears in GIXRD). **c, d,** The back-side shows negligible amorphization after the impact, indicating preservation of the crystalline structure.

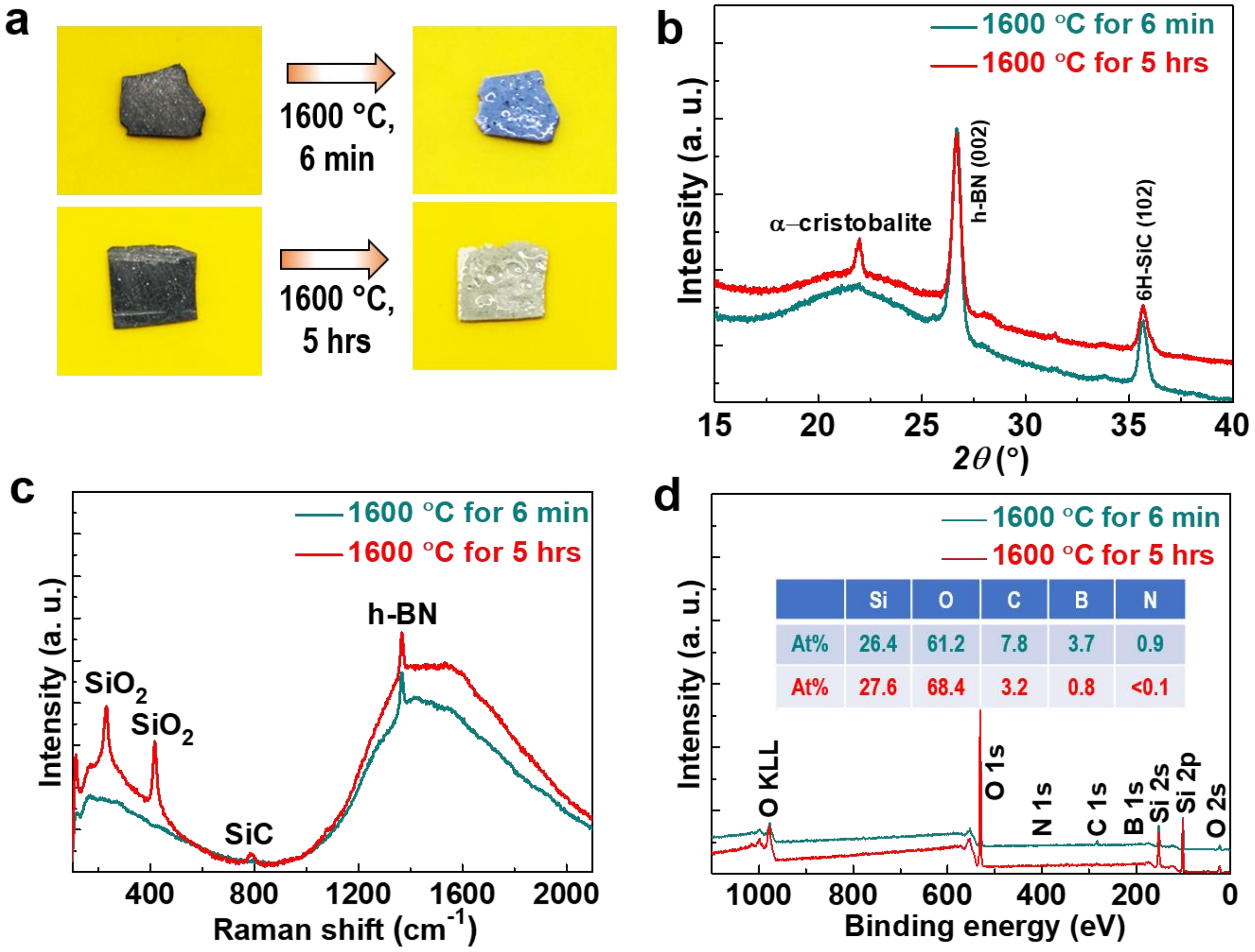

**Fig. S25: Characterizations of the composite after furnace oxidation at 1600 °C. a,** Images of the sample annealed in a box furnace at 1600 °C for 6 min, approximating the HVOF impact duration, and for 5 h to assess longer-term oxidation behavior. **b, c, d,** XRD, Raman spectroscopy, and XPS reveal the formation of $SiO_2$, comprising both amorphous and *α*-cristobalite phases, following high-temperature annealing in air.

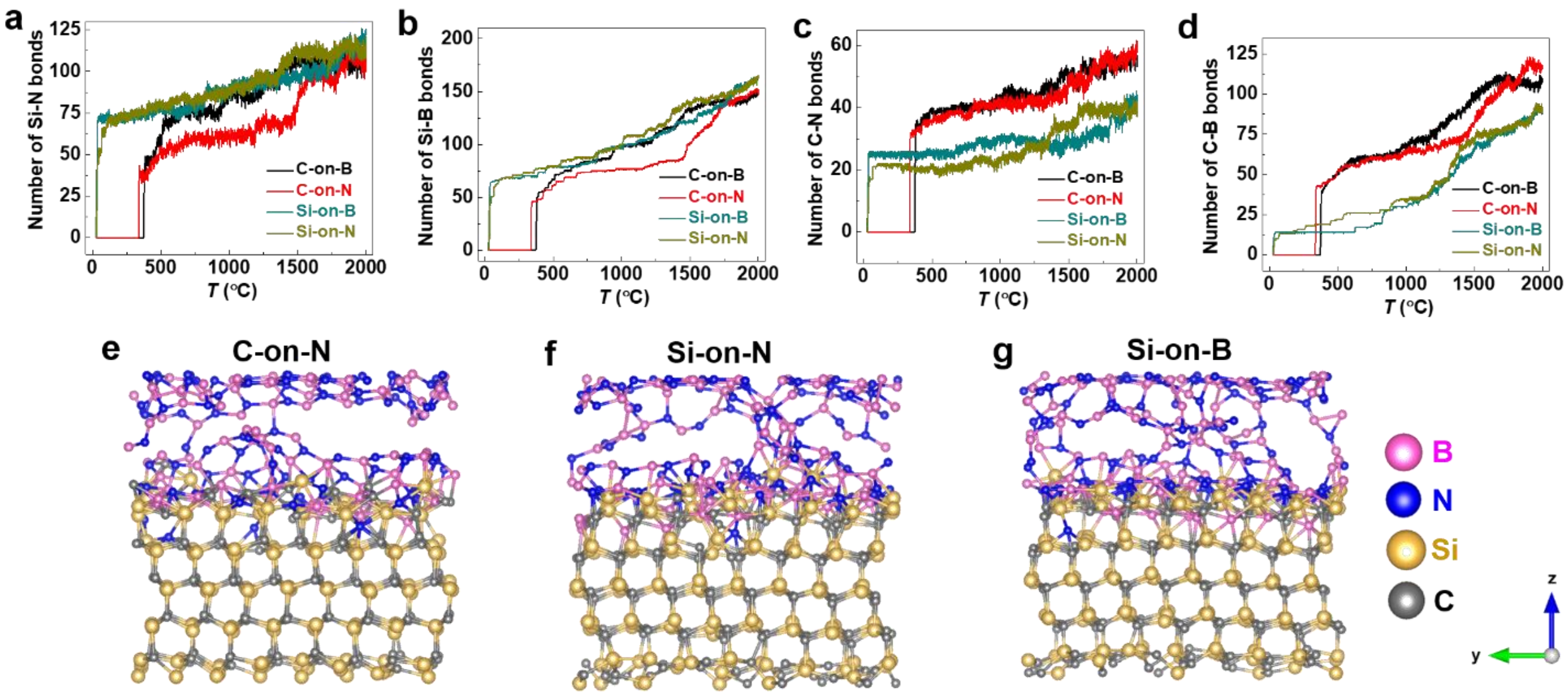

**Fig. S26: Temperature-dependent interfacial bond evolution during continuous heating**: **a,** Si-N, **b,** Si-B, **c,** C-N, and **d,** C-B bonds for C-on-B, C-on-N, Si-on-B, and Si-on-N interface configurations. Side-view atomic configurations of interfaces, at 1500 °C for **e,** C-on-N, **f,** Si-on-N, **g,** Si-on-B, respectively.

**Table S5: Properties of sintered 6H-SiC, h-BN and SiC-BN ceramic composite.**

| Ceramic | Density ($g/cm^3$) | Hardness (GPa) | Fracture toughness ($MPa \cdot m^{1/2}$) | Elastic modulus (GPa) | Yield strength (GPa) | Thermal conductivity (at 1500 °C) (W/mK) | HVOF impact test |
|---|---|---|---|---|---|---|---|
| **SiC** | 1.78 | 28.8# | 0.77-2# | 227.4# | 7.2# | ~18.10* | Shatters in 3-4 s |
| **h-BN** | 1.97 | 0.43 | 1.52 | 27.2 | 0.245± | ~41.79 | Erodes after few min |
| **SiC-BN** | 1.85 | 1.39 | 2.02 | 37.5 | 0.861 | ~11.57 | Stable for > 6 min |

#From D. Zhang *et al.*,[38]

*Extrapolating the plot from S. N. Perevislov *et al.*,[30]

±From A. Biswas *et al.*,[13]

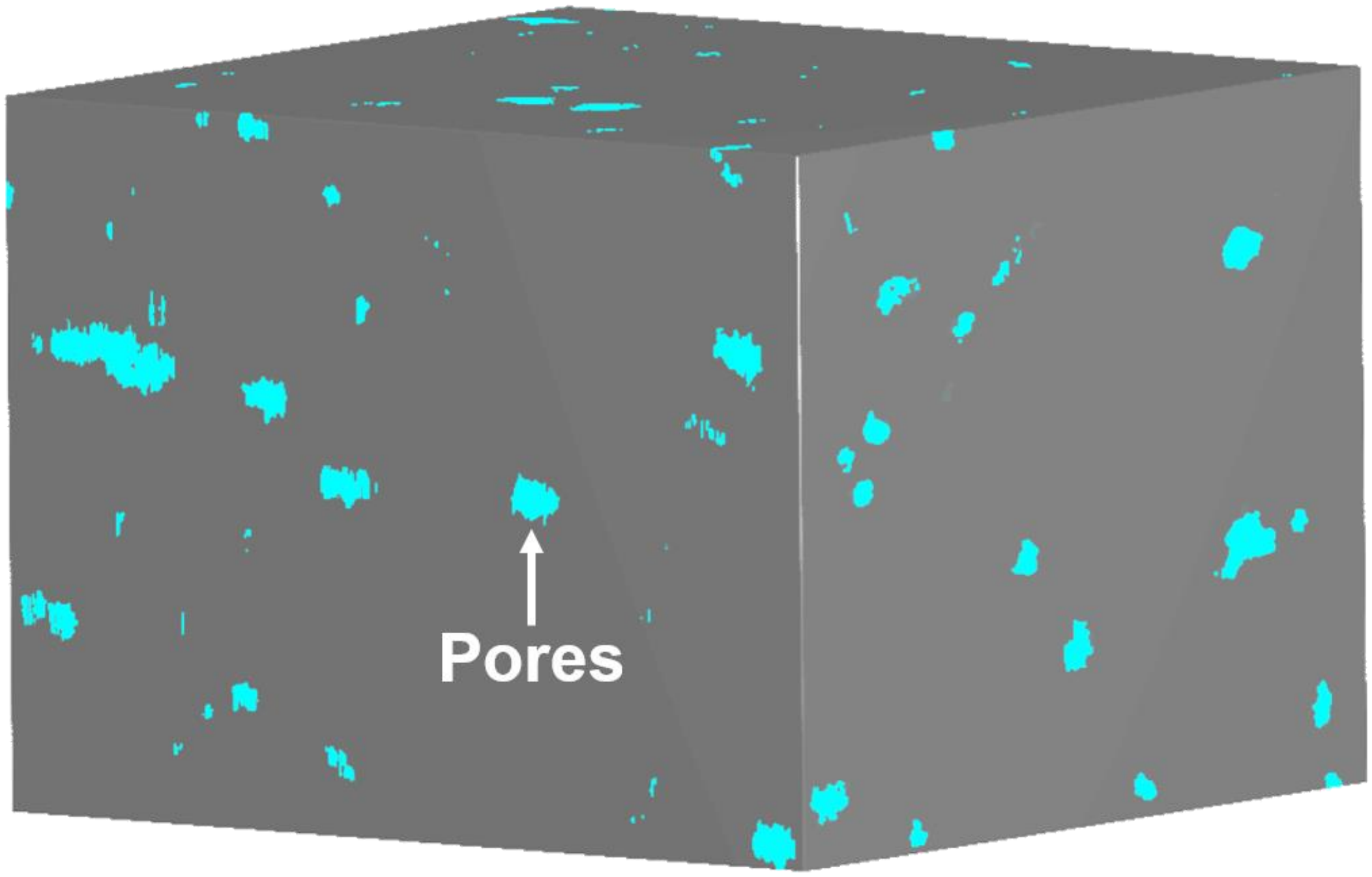

**Fig. S27: Pores in the composite after the HVOF exposure.** X-ray microscopy (XRM) image reveal the presence of pores in the bulk even after the HVOF exposure. (cyan: pores, grey: SiC-BN). Box dimensions: $1.21 \times 1.08 \times 0.87\ mm^3$.

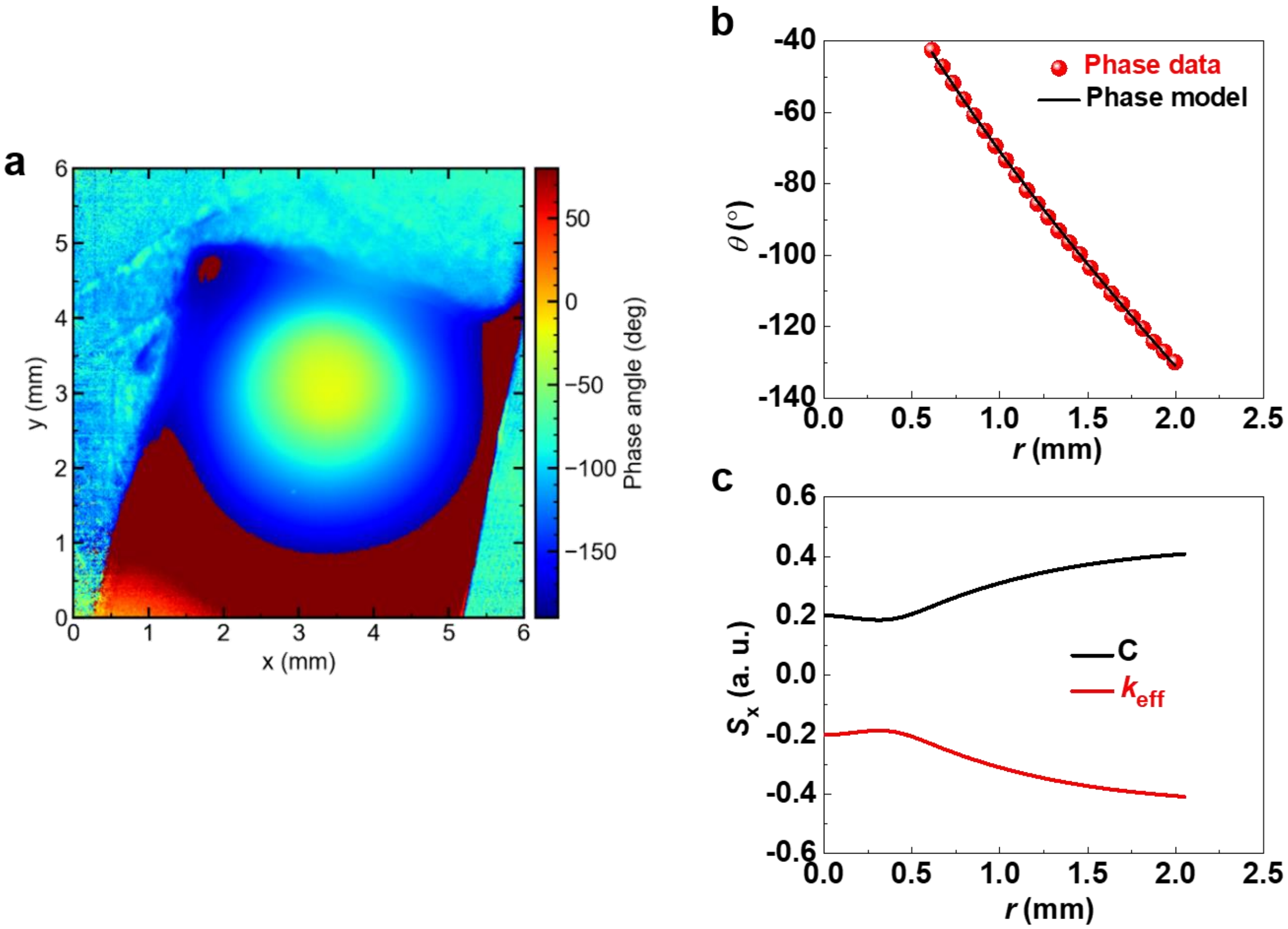

**Fig. S28: A representative LIT measurement and the associated sensitivity analysis used to extract $\boldsymbol{k_{eff}}$. a,** Two-dimensional phase-angle map at 25 °C obtained from the IR-camera measurement in LIT setup, showing the spatial phase response around the modulated laser-heating spot. **b,** Measured radial phase response (symbols) and corresponding thermal-model fit (solid line) used to extract the effective thermal conductivity. **c,** Radial sensitivity coefficients of the modeled phase response to volumetric heat capacity, $C$, and effective thermal conductivity, $k_{eff}$.

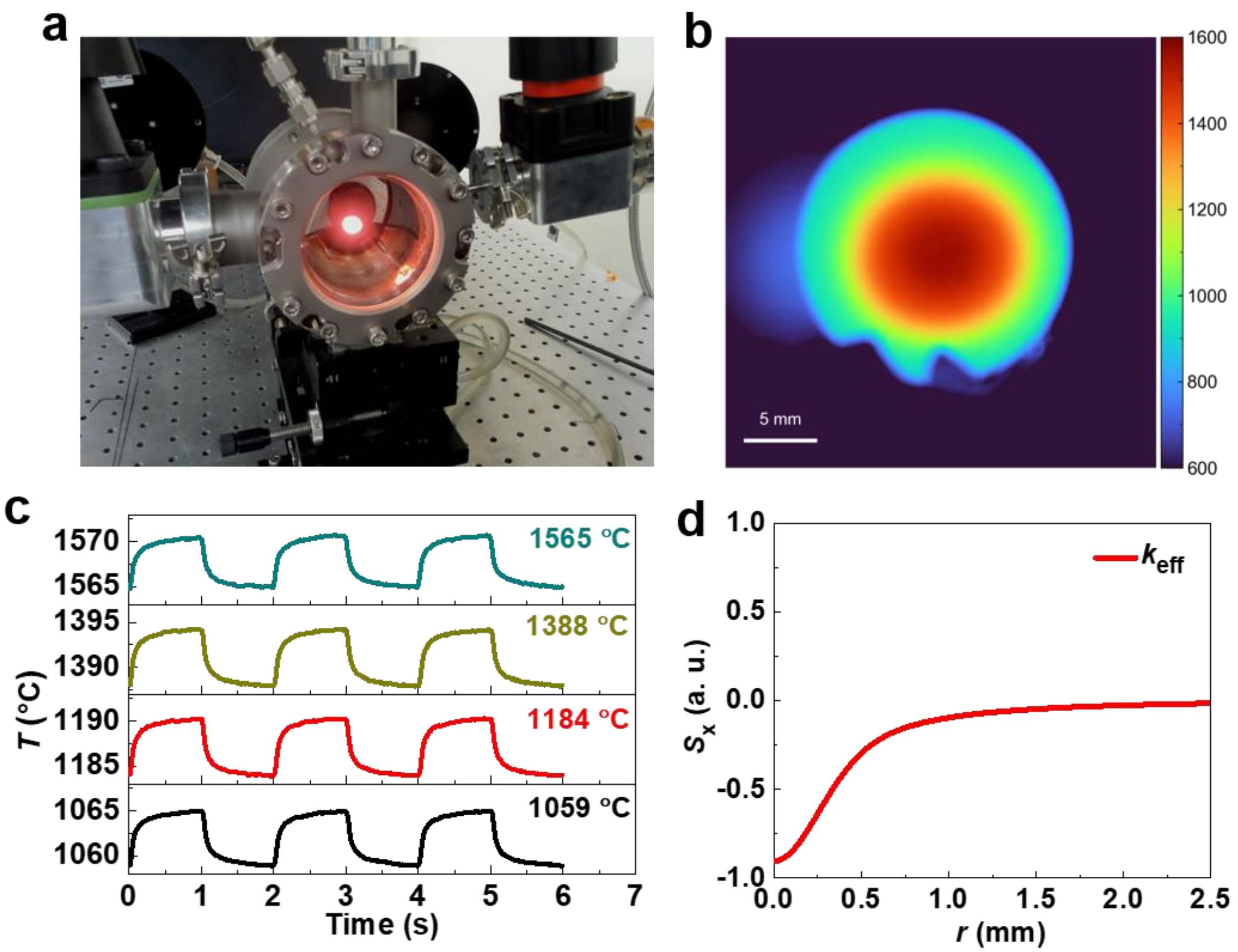

**Fig. S29: The SSTDR measurement and analysis ≥1000 °C**. **a,** Photograph of the SSTDR experimental chamber during laser heating of the SiC-BN. **b,** Representative IR-camera image showing the surface-temperature distribution during high-temperature laser heating. **c,** Temporal temperature response during application of the 532 nm perturbation laser at different background temperatures. The step increase corresponds to the small differential temperature rise, $\Delta T$, used for the SSTDR analysis. **d,** Radial sensitivity coefficient of the modeled temperature response to the effective thermal conductivity over the analysis region.

**Table S6.** WDS analytical conditions and primary standards for SiC-BN point analysis (10 kV, 100 nA, focused beam ~0 nm).

| # | Element | X-ray | Crystal | CH | Acc. V (kV) | Peak pos. (mm) | BG – (mm) | BG + (mm) | Peak (s) | BG (s) | Mode | Gain | HV | Standard | Std. mass % |
|---|---|---|---|---|---|---|---|---|---|---|---|---|---|---|---|
| 1 | C | $K_\alpha$ | LDE2 | 1 | 10 | 127.381 | 4.47 | 5 | 10 | 5 | Dif | 64 | 1700 | C_graphite_SPI | 99.99 |
| 2 | B | $K_\alpha$ | LDE2 | 1 | 10 | 195.213 | 5 | 5 | 10 | 5 | Dif | 64 | 1700 | BN_SPI | 43.60 |
| 3 | B | $K_\alpha$ | LDEB | 1 | 10 | (128.853[1]) | 6 | 6 | 10 | 5 | Dif | 128 | 1758 | Cal-STD | 100 |
| 4 | N | $K_\alpha$ | LDE1 | 4 | 10 | 147.882 | 7 | 7 | 10 | 5 | Dif | 64 | 1654 | BN_SPI | 56.44 |
| 5 | O | $K_\alpha$ | LDE1 | 4 | 10 | 110.230 | 5 | 5 | 10 | 5 | Dif | 64 | 1654 | Quartz_SPI | 53.255 |
| 6 | Si | $K_\alpha$ | TAP | 4 | 10 | 77.300 | 6 | 6 | 10 | 5 | Int | 32 | 1678 | Quartz_SPI | 46.745 |

[1] Peak position from WDS map acquisition (LDEB crystal, CH1). The point-analysis LDEB entry (row 3) lists peak position as 0 mm in the raw data file, indicating the crystal was used in the mapping recipe; the map-derived position (128.853 mm) is reported here for reference. Measurement order: Acquisition 1: C $K_\alpha$ (CH1/LDE2), N $K_\alpha$ (CH4/LDE1); Acquisition 2: B $K_\alpha$ (CH1/LDE2), O $K_\alpha$ (CH4/LDE1); Acquisition 3: B $K_\alpha$ (CH1/LDEB), Si $K_\alpha$ (CH4/TAP). CH2, CH3, and CH5 are not used in this recipe. Matrix correction: CITZAF/PAP.

**Table S7.** EPMA WDS map acquisition parameters.

| Parameter | Value |
|---|---|
| Accelerating voltage | 10 kV |
| Beam current | ~50 nA (probe current $5.021 \times 10^{-8}$ A) |
| Beam mode | SPOT (focused, ~0 µm probe diameter) |
| Dwell time per pixel | 20 ms |
| Map dimensions | 650 × 500 pixels |
| Pixel spacing | 0.60 × 0.60 µm |
| Map area | 390 × 300 µm |
| Stage position | X = −3.574 mm, Y = −14.354 mm, Z = 12.167 mm |
| Elements mapped (CH1) | B Kα (LDEB crystal, peak pos. 128.853 mm) |
| Elements mapped (CH4) | N $K_\alpha$ (LDE1), O $K_\alpha$ (LDE1), Si $K_\alpha$ (TAP) |
| Elements mapped (CH1, Acq. 2) | C $K_\alpha$ (LDE2) |
| Crystals / channels used | CH1: LDEB (B), LDE2 (C); CH4: LDE1 (N, O), TAP (Si) |
| Matrix correction | CITZAF / PAP |

**Note:** The map was acquired at 10 kV to minimize the electron interaction volume and beam spread in the SiC-BN composite. The low accelerating voltage is appropriate for light-element (Z ≤14) characterization but limits excitation of heavier elements. Concentration images show mass% as calibrated against the same standards used for point analysis.

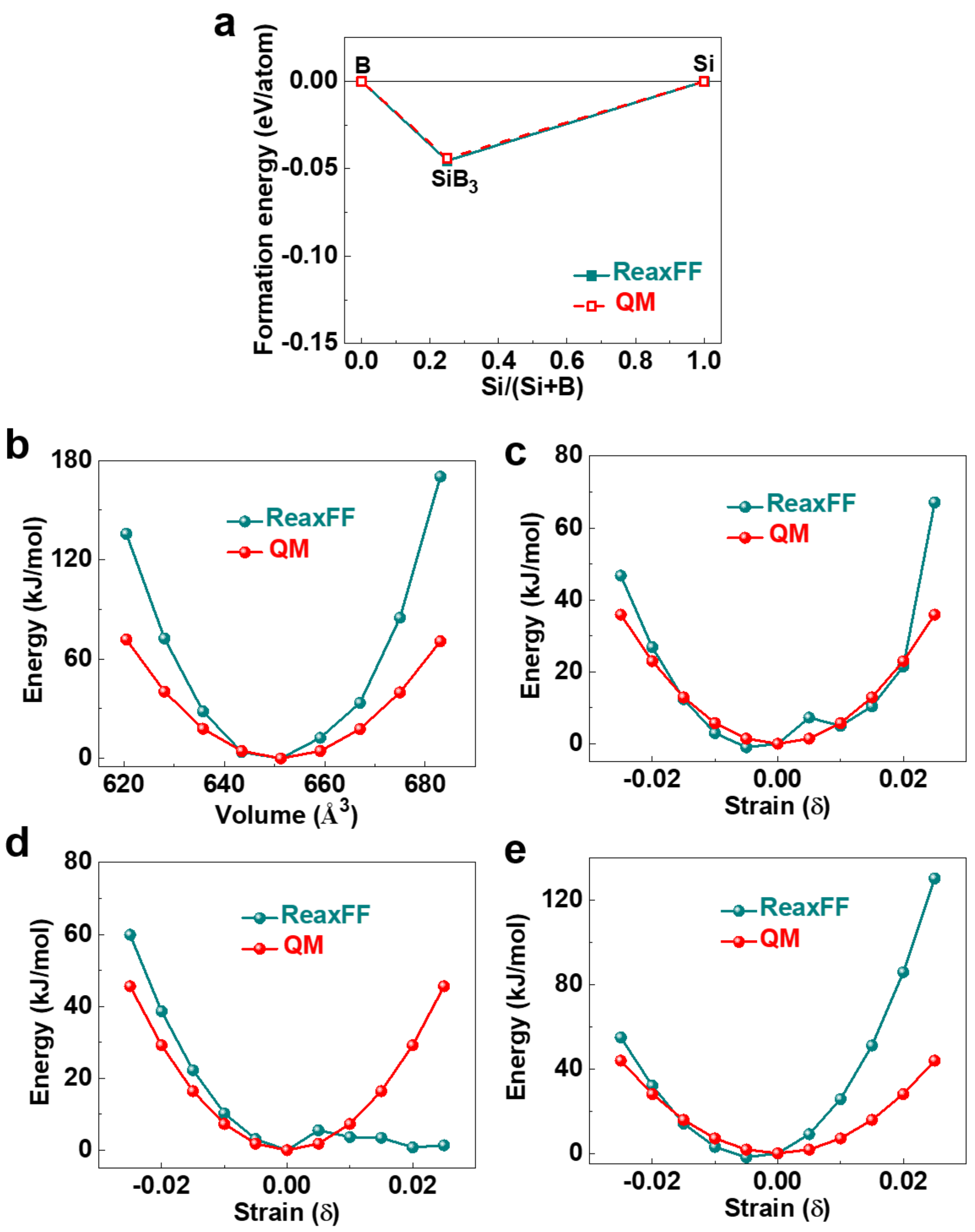

**Fig. S30: Si-B training set and Overall strain-dependent trends. a,** Computed heat of formation vs composition curves for stable stoichiometric silicon boride compounds. **b-e,** ReaxFF fit to equations of state (EOS) from QM calculations, corresponding to **b,** bulk modulus, **c,** $C_{11}$, **d,** $C_{22}$, and **e,** $C_{33}$.

**Table S8.** Finite element analysis (FEA) inputs and results at 1500 °C (medium mesh).

| Quantity | Value | Basis |
|---|---|---|
| $k_{SiC\text{-}BN}$ | 11.57 ± 1.67 W/mK | Measured; interpolated from 1386-1564 °C |
| $k_{SiC}$ | 18.10 W/mK | Hot-pressed SiC; extrapolated beyond 1430 °C |
| $k_{BN}$ | 41.79 W/mK | SPS h-BN; extrapolated between 1398-1507 °C |
| SiC/BN resistance (R) | $2.06 \times 10^{-8}$ m²K/W (±1σ: 1.25–3.5 × $10^{-8}$) | Calibrated |
| $k_{eff}$, calibrated interfaces | 11.58 W/mK | FEA |
| $k_{eff}$, perfect interfaces (R = 0) | 24.51 W/mK | FEA |
| $k_{eff}$, R = 0 and pores filled | 28.01 W/mK | FEA |
| Interface knock-down | 2.12× | Ratio of the two R cases |

## Supplementary Note

### Additional discussion on thermal conductivity

The effective conductivity is controlled not only by the intrinsic conductivities of SiC and h-BN, but also by their connectivity, the interfaces, grain boundaries, intrinsic pores, and local constrictions. When these comparatively weakly temperature-dependent resistances constitute a substantial fraction of the overall resistance to heat flow, even pronounced changes in intrinsic phonon scattering within the constituent phases can produce only a modest change in the measured $k_{eff}$.

The low- and high-temperature measurements provide additional insight into how this balance evolves with temperature. In the LIT regime, $k_{eff}$ increases only modestly at lower temperatures and then approaches a nearly constant value through 600 °C. Physically, the increase in lattice heat capacity over this temperature range tends to increase thermal conductivity. At the same time, heat transport is already strongly constrained by defects, grain boundaries, interfaces, pores, and constricted heat-flow pathways, so the increasing phonon-phonon scattering has a comparatively smaller influence on the overall transport and can partially offset the increase in heat capacity. The modest increase in $k_{eff}$ at lower temperatures is therefore consistent with the increase in lattice heat capacity while the dominant microstructural scattering processes remain comparatively weakly temperature dependent.

At high temperatures, the SSTDR values show that $k_{eff}$ remains nearly unchanged from 1000 to 1600 °C. By these temperatures, the lattice heat capacity is expected to be much closer to its high-temperature limit and therefore changes only weakly with further increases in temperature. With the lattice heat capacity approaching its high-temperature limit and heat transport remaining strongly constrained by comparatively temperature-insensitive microstructural resistances, the nearly temperature-independent $k_{eff}$ observed by SSTDR is consistent with this transport picture. The persistence of this behavior to ~1600 °C suggests that interfaces, grain boundaries, pores, and other structural features continue to strongly limit the macroscopic thermal response even under ultrahigh-temperature conditions.

## Supplementary Videos

**Supplementary Video 1: HVOF torch impact test on the SPS SiC-BN composite**. High-velocity oxygen fuel (HVOF) torch impact test on the sintered SiC-BN composite (continues torch-in duration of 6 min). The distance between the torch and the sample was 20 cm, producing a front-face surface temperature of 1500 °C.

**Supplementary Video 2: HVOF torch impact test on the SPS SiC disk**. High-velocity oxygen fuel (HVOF) torch impact test on the SPS sintered 6H-SiC that shatters in 3-4 s. The distance between the torch and the sample was 20 cm.

**Supplementary Video 3: HVOF torch impact test on the commercial SiC slab**. High-velocity oxygen fuel (HVOF) torch impact test on the commercially available SiC slab that shatters in 2-3 s. The distance between the torch and the sample was 20 cm.

**Supplementary Video 4: HVOF torch impact test on the SPS h-BN disk**. High-velocity oxygen fuel (HVOF) torch impact test on the SPS h-BN disk. The disk starts eroding after 2-3 min. The distance between the torch and the sample was 20 cm.

**Supplementary Video 5: HVOF torch impact thermal cycling test on the SPS SiC-BN disk**. High-velocity oxygen fuel (HVOF) torch impact thermal cycling test (3 cycles with a torch-in duration of 6 min in the heating part of the cycle and 10 s of torch-off time between each cycle) on the SiC-BN composite, showing its thermal-shock resistance.

**Supplementary Video 6: HVOF torch impact test on the SPS SiC-BN composite at 2000 °C**. High-velocity oxygen fuel (HVOF) torch impact test showing that the composite disk shatters after 20 s. The stand-off torch-to-sample distance was reduced to 15 cm, to produce the front-face surface temperature of 2000 °C.